\documentclass[dvipsnames]{aa}  

\usepackage{graphicx}
\usepackage{subfigure}
\usepackage{txfonts}
\usepackage{amssymb}
\usepackage{amsmath}
\usepackage{natbib}
\usepackage{xcolor}
\usepackage{url}
\usepackage{placeins}
\usepackage{rotating}
\usepackage{float}
\DeclareMathAlphabet{\mathitbf}{OML}{cmm}{b}{it}
\newcommand{\Avec}{\mathitbf{A}}
\newcommand{\Apvec}{\mathitbf{A}_{\mathrm{0}}}
\newcommand{\Bvec}{\mathitbf{B}}

\newcommand{\Bz}{\mathit{B_z}}
\newcommand{\Bjvec}{\mathitbf{B}_{\rm J}}
\newcommand{\Bpvec}{\mathitbf{B}_{\mathrm{0}}}

\newcommand{\Ef}{E_{\rm F}}
\newcommand{\Etot}{E}

\newcommand{\Ediv}{E_{\rm div}}
\newcommand{\Edivprime}{E_{\rm div}/E}
\newcommand{\Hj}{H_{\rm J}}
\newcommand{\Hv}{H_{\rm V}}

\newcommand{\mxmx}{{\rm Mx}^2}
\newcommand{\mx}{{\rm Mx}}

\newcommand{\erg}{{\rm erg}}
\newcommand{\kmps}{{$\mathrm{~km~s}^{-1}$}}

\newcommand{\Rsun}{{$~\mathrm{R}_{\odot}$}}

\begin{document}

\title{The 2019 International Women's Day event }
    \subtitle{A two-step solar flare with multiple eruptive signatures and low Earth impact}
\author{M. Dumbovi\'c\inst{1}
          \and A.M. Veronig\inst{2, 3}
          \and T. Podladchikova\inst{4}
          \and J.K. Thalmann\inst{2}
          \and G. Chikunova\inst{4}
          \and K. Dissauer\inst{5}
          \and J. Magdaleni\'c\inst{6, 7}
          \and M. Temmer\inst{2}
          \and J. Guo\inst{8}
          \and E. Samara\inst{6, 7}
          }

\institute{Hvar Observatory, Faculty of Geodesy, University of Zagreb, Kaciceva 26, HR-10000 Zagreb, Croatia\\
        \email{mdumbovic@geof.hr}\\
         \and  University of Graz, Institute of Physics, Universit\''atsplatz 5, 8010 Graz, Austria\\
         \and University of Graz, Kanzelh\''ohe Observatory for Solar and Environmental Research, Kanzelh\''ohe 19, 9521 Treffen, Austria\\
         \and Skolkovo Institute of Science and Technology, Bolshoy Boulevard 30, bld. 1, Moscow 121205, Russia\\
          \and NorthWest Research Associates, 3380 Mitchell Lane, Boulder, CO 80301, USA\\
         \and Solar-Terrestrial Centre of Excellence -- SIDC, Royal Observatory of Belgium, 3 Avenue Circulaire, 1180 Uccle, Belgium\\
         \and Center for mathematical Plasma Astrophysics, Department of Mathematics, KU Leuven, Celestijnenlaan 200B, B-3001 Leuven, Belgium\\
         \and School of Earth and Space Sciences, University of Science and Technology of China, JinZhai Rd. 96, 230026, Hefei, PR China\\
         }
        
\abstract
{We present a detailed analysis of an eruptive event that occurred on early 2019~March~8 in active region AR~12734, to which we refer as the International Women's day event. The event under study is intriguing in several aspects:
1) low-coronal eruptive signatures come in ``pairs'' (i.e.\, there is a double-peak flare, two coronal dimmings, and two EUV waves); 2) although the event is characterized by a complete chain of eruptive signatures, the corresponding coronagraphic signatures are weak; 3) although the source region of the eruption is located close to the center of the solar disc and the eruption is thus presumably Earth-directed, heliospheric signatures are very weak with little Earth-impact.}
{In order to understand the initiation and evolution of this particular event, we perform a comprehensive analysis using a combined observational-modeling approach.}
{We analyze a number of multi-spacecraft and multi-instrument (both remote-sensing and in\,situ~) observations, including Soft X-ray, (extreme-) ultraviolet (E)UV), radio and white-light emission, as well as plasma, magnetic field and particle measurements. We employ 3D nonlinear force-free (NLFF) modeling to investigate the coronal magnetic field configuration in and around the active region, the graduated cylindrical shell (GCS) model to make a 3D reconstruction of the CME geometry and the 3D magnetohydrodynamical (MHD) numerical model EUHFORIA (EUropean Heliospheric FORecasting Information Asset) to model the background state of the heliosphere.}
{Our results reveal a two-stage C1.3 flare, associated with two EUV waves that occur in close succession and two-stage coronal dimmings that evolve co-temporal with the flare and type II and III radio bursts. Despite its small GOES class, a clear drop in magnetic free energy and helicity is observed during the flare. White light observations do not unambiguously indicate two separate CMEs but rather a single entity most likely composed of two sheared and twisted structures corresponding to the two eruptions observed in the low corona. The corresponding interplanetary signatures are that of a small flux rope (SFR) with indications of strong interactions with the ambient plasma which result in negligible geomagnetic impact.}
{Our results indicate two subsequent eruptions of two systems of sheared and twisted magnetic fields, which merge already in the upper corona and start to evolve further out as a single entity. The large-scale magnetic field significantly influences both, the early and the interplanetary evolution of the structure. During the first eruption the stability of the overlying field was disrupted which enabled the second eruption. We find that during the propagation in the interplanetary space the large-scale magnetic field, i.e.\,, the location of heliospheric current sheet (HCS) between the AR and the Earth likely influences propagation and the evolution of the erupted structure(s).}
\keywords{Sun: coronal mass ejections (CMEs) -- Sun: flares -- Sun: magnetic fields}
\maketitle
\clearpage
%
\section{Introduction}
\label{intro}

Solar flares and coronal mass ejections (CMEs) are the most powerful events in our solar system. They come in various sizes and flavors, and are a direct manifestation of instabilities of the strong magnetic fields concentrated in sunspots and active regions. The current paradigm is that both may be related to the eruption of a sheared magnetic arcade or a flux rope \citep{patsourakos2020}, i.e.\, a magnetic structure with field lines helicoidally winding around a central axis. In support, CMEs are often observed to be filled with helical magnetic field, indicative of a magnetic flux rope being carried away from the Sun \citep[e.g.,][]{2011LRSP....8....1C}. As the flux rope erupts, free magnetic energy that was built up is impulsively released and the surrounding magnetic field rearranged \citep{Priest2002,Fletcher2011,Benz2017,Green2018,2021arXiv210404261T}.

CMEs may be accompanied by large-scale propagating disturbances, EUV waves, which are most commonly interpreted as coronal large-amplitude fast-mode MHD waves or shocks, initiated by the impulsive lateral expansion of the CME flanks \citep[e.g.][]{Veronig2008,Warmuth2015,long17a,Veronig2018}, although other interpretations do exist \citep[e.g.][]{chen11b,chandra18}. A subset of EUV waves is accompanied by type II radio bursts, i.e.\, slowly drifting bands of radio emission observed in dynamic spectra, which are signatures of shock waves traveling outwards through the corona and  interplanetary space \citep{Warmuth2015}, and are closely related to CMEs \citep{Vrsnak2008}. Another relatively common low coronal signature of the eruption are coronal dimmings, which  are reflections of the expansion of coronal fields and mass depletion associated with CMEs \citep{hudson1996,thompson1998,dissauer2018b,dissauer2019,chikunova2020}. It has been shown that the localized core dimming regions tend to be rooted at the footpoints of the ejected flux rope \citep{Sterling1997,Webb2000,2017SoPh..292...93T,Veronig2019}, and that the spatio-temporal evolution of the dimming regions provides insight into the initiation and early evolution of the eruption \citep{Miklenic2011,Prasad2020}, as well as into the interaction of the flux rope footpoints with the surrounding magnetic field \citep[e.g.][]{lorincik21}.

Flare properties can outline footprints of a flux rope \citep[e.g.][]{janvier14}. Although there seems to be no tendency showing that large flares are preferentially associated with EUV waves \citep{,nitta2013,muhr2014,long17a}, there is a distinct relation between the flare reconnection fluxes and magnetic fluxes covered by dimming regions \citep{dissauer2018b}. On the other hand, there is no one-to-one relation between CMEs and flares. The so-called stealth CMEs can occur without solar flares, or any other low coronal signatures \citep{robbrecht09,ma10,nitta2017} and so-called confined flares are not accompanied by CMEs \citep{svestka86, schmieder15,li2020}. There seems to be a tendency that large flares are associated with CMEs, whereas small flares are confined. As shown in \cite{Yashiro2006}, the CME-association is a  steeply increasing function of flare importance class, with $>$90\% of X-class flares accompanied by a CME, while only about $10\%$ of C1 class flares are associated with a CME. However, there exist also notable exceptions to this general trend, such as active region (AR) 12192, which was the source of 6 X- and 30 M-class flares during October 2014, out of which only one flare was associated to a CME \citep{Thalmann2015,Sun2015,Veronig2015}.

Nevertheless, there seems to be a general trend that more energetic CMEs occur in close association with powerful flares and the restructuring of the overlying fields is ultimately related to various wave and wave-like signatures in the low corona \citep[see also][and references therein]{hudson11}. In addition, due to the feedback relationship between the CME dynamics and the reconnection process, such CMEs would tend to be more massive, and thus more likely to have pronounced white light signatures in the upper corona and interplanetary space \citep{colaninno09,howard09}. Finally, when such CMEs have source region close to the central meridian, they are more likely to hit the near-Earth spacecraft head-on, and thus tend to produce pronounced magnetic cloud signatures, i.e.\, a low-temperature structure with smoothly rotating enhanced magnetic field and linearly decreasing plasma speed \citep[e.g.][and references therein]{kilpua17}. Magnetic clouds tend to produce the strongest geomagnetic effects, as they are more likely to carry a long-lasting enhanced south oriented magnetic field needed to produce a geomagnetic storm \citep{kilpua17b}. Indeed, statistical studies have shown that fast CMEs associated with large flares with source regions close to the central meridian are more likely to be more geo-effective \citep{srivastava04,dumbovic15,scolini18}, as well as CMEs which host strong magnetic flux from  dimming areas \citep{chertok13}. In addition, highly geo-effective CMEs are often associated with strong interplanetary type II radio bursts with low end frequency \citep{vasanth15}. However, we note that these general trends do not mean that all strong flares and energetic CMEs will result in extreme space weather events. As already discussed, and also shown by \citet{schmieder20}, exceptions to this generalization do exist.

In this paper, we study an intriguing event which seems to oppose several of the general tendencies described above, a small double-peaked C1.3 flare that occurred under solar minimum conditions on 2019~March~8 (SOL-2019-03-08T03:07C1.3), the International Woman's Day event. Despite its low GOES class, the flare was associated with a plethora of phenomena typical for solar eruptions, including coronal dimmings, two EUV waves, type II and III bursts and two coronal loop ejections, which come in ``pairs''. Although the event is characterized by very pronounced low coronal signatures, the observed white light signatures in the upper corona and interplanetary space are quite weak. This is perhaps not uncommon for a small flare, but is surprising given that the event has a complete chain of eruptive signatures, which, according to some general tendencies, should indicate a strong eruption. Moreover, the event had a very low impact on the near-Earth environment, although it was Earth-directed. We study in detail the initiation and early evolution of the event, with a particular focus on the two-step process revealed in both, the flare energy release and the different eruption signatures. Finally, we follow its faint traces into interplanetary space and investigate its potential in\,situ~ detection and effects at 1 AU.

\begin{figure*}
\centering
\includegraphics[width=0.85\textwidth]{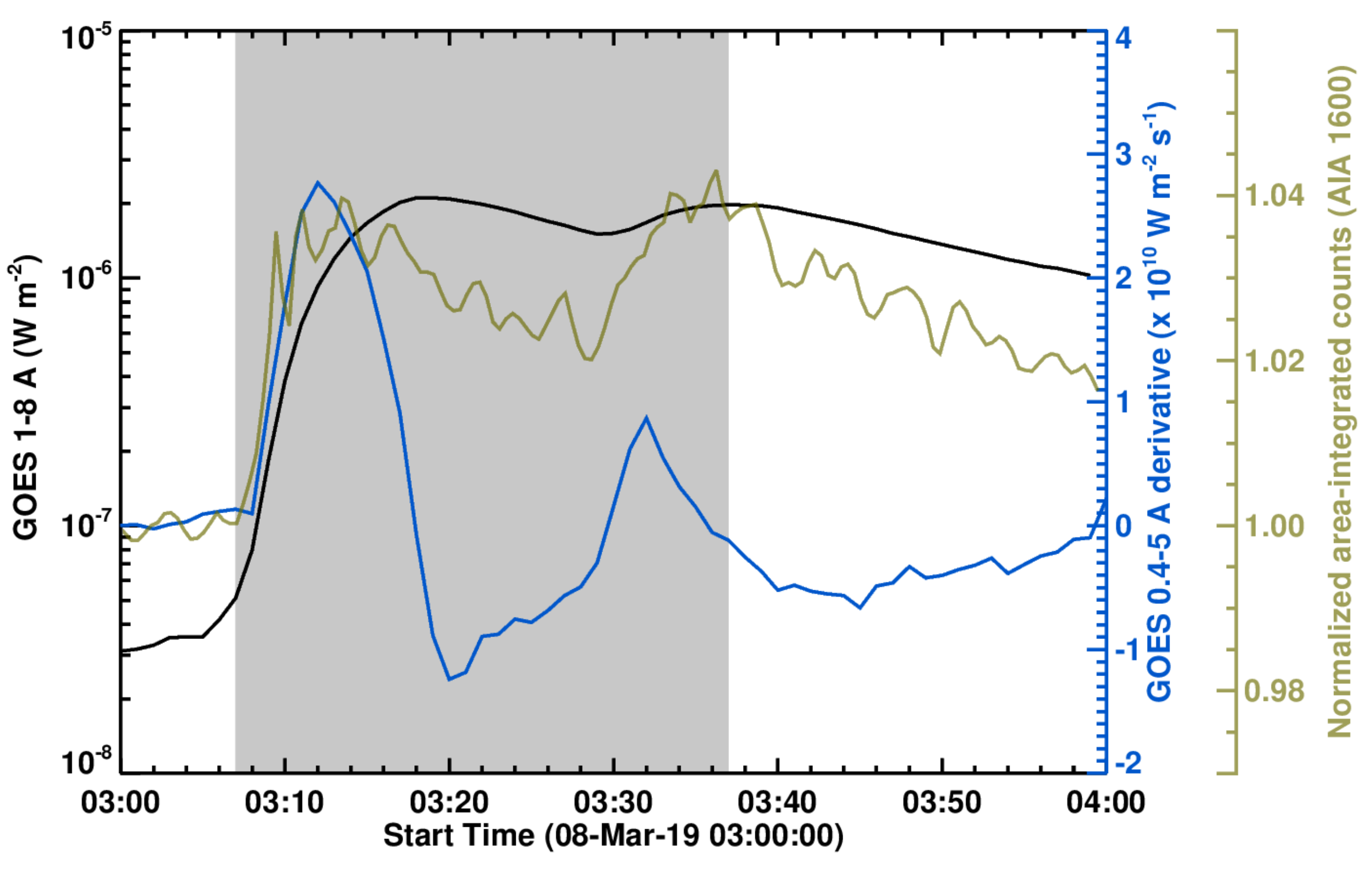}
\caption{Soft X-ray and ultraviolet emission during the C1.3 flare (SOL-2019-03-08T03:07) on 2019~March~8. We show the full-disk integrated GOES 1--8~\AA\ SXR emission (black curve), the time derivative of the GOES 0.5--4~\AA\ emission (blue curve), and the NOAA AR~12734 area-integrated AIA 1600~\AA\  emission (green) normalized to the pre-flare state at 03:00:23~UT. The gray-shaded vertical band marks the  extended impulsive phase of the C1.3 flare, covering the time span between the nominal GOES start time and the time of the second peak in the GOES 1--8~\AA\ SXR emission.}
\label{fig1}
\end{figure*}

\begin{figure*}
\centering
\includegraphics[width=0.83\textwidth]{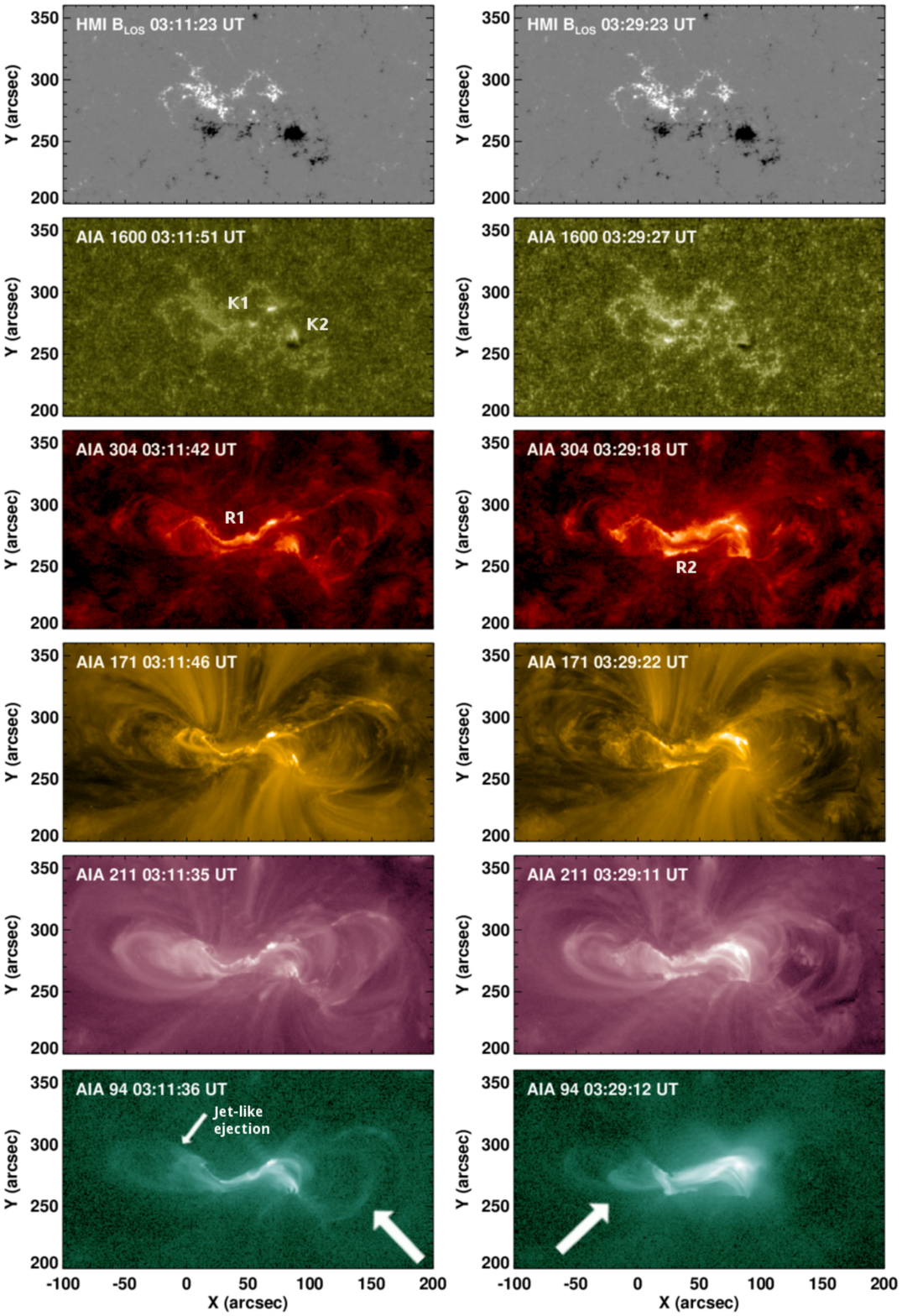}
\caption{ Magnetic field and (E)UV emission during the extended impulsive phase of the C1.3 flare on 2019~March~8 (SOL-2019-03-08T03:07C1.3). The panels in the two columns correspond to the two phases of the double peak flare and two eruptions from the same NOAA AR 12734. Namely, snapshots are shown at two time steps corresponding to the first (around 03:12~UT; {\it left column}) and second (around 03:30~UT; {\it right column}) flare-related eruption. Top to bottom rows show HMI line-of-sight magnetic field maps, AIA 1600, 304, 171, 211, and 94~\AA\ filtergrams. Labels ``K1'' and ``K2'' in AIA 1600~\AA\ filtergrams refer to locations of flare kernels; labels ``R1'' and `R2'' in AIA 304~\AA\ filtergrams to corresponding flare ribbons. Large white arrows in the AIA 94~\AA\ filtergrams point to the location of erupting loops during stage one and two of the flare process, respectively. The small arrow in the bottom left AIA 94~\AA\ filtergram points to the location of a narrow ejection to the north-east direction during the first stage. A movie accompanying the figure is available in the electronic material (movie1).}
\label{fig2}
\end{figure*}

\begin{figure*}
\centering
\includegraphics[width=1.0\textwidth]{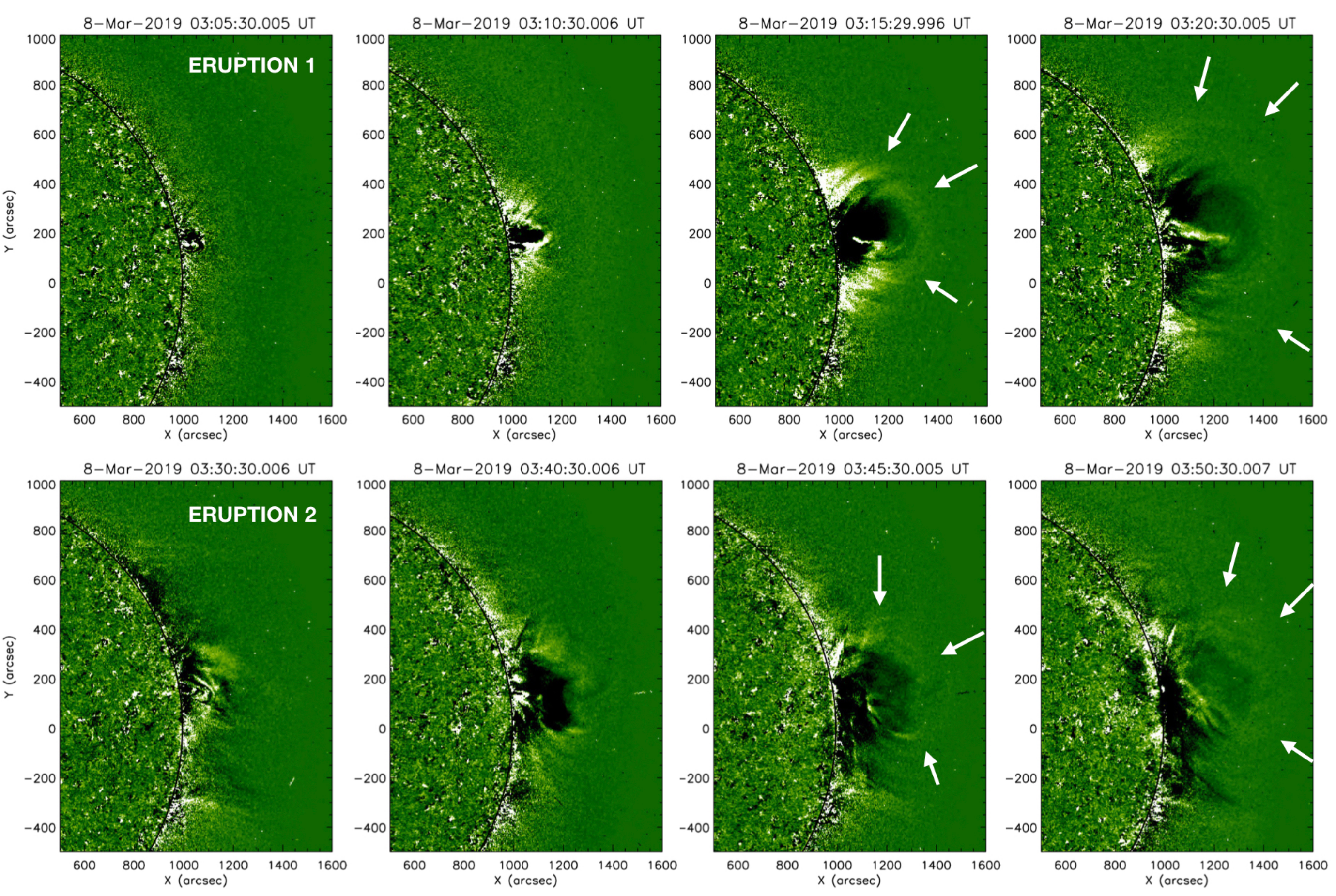}
\caption{STEREO-A/EUVI 195~\AA\ running difference images showing the early evolution of the two eruptions on the Western limb. Top row: First ejection which is initiated low in the corona and associated with the first peak in the GOES/SXR flux at around 03:18~UT. Bottom row: Second ejection, starting higher up in the corona corresponding to the second peak in GOES/SXR at around 03:37~UT. The white arrows indicate the leading edge of the eruptions. A movie accompanying the figure is available in the electronic material (movie2).}
\label{fig3}
\end{figure*}

\begin{figure*}
\begin{subfigure}
\centering
\includegraphics[width=1.0\textwidth]{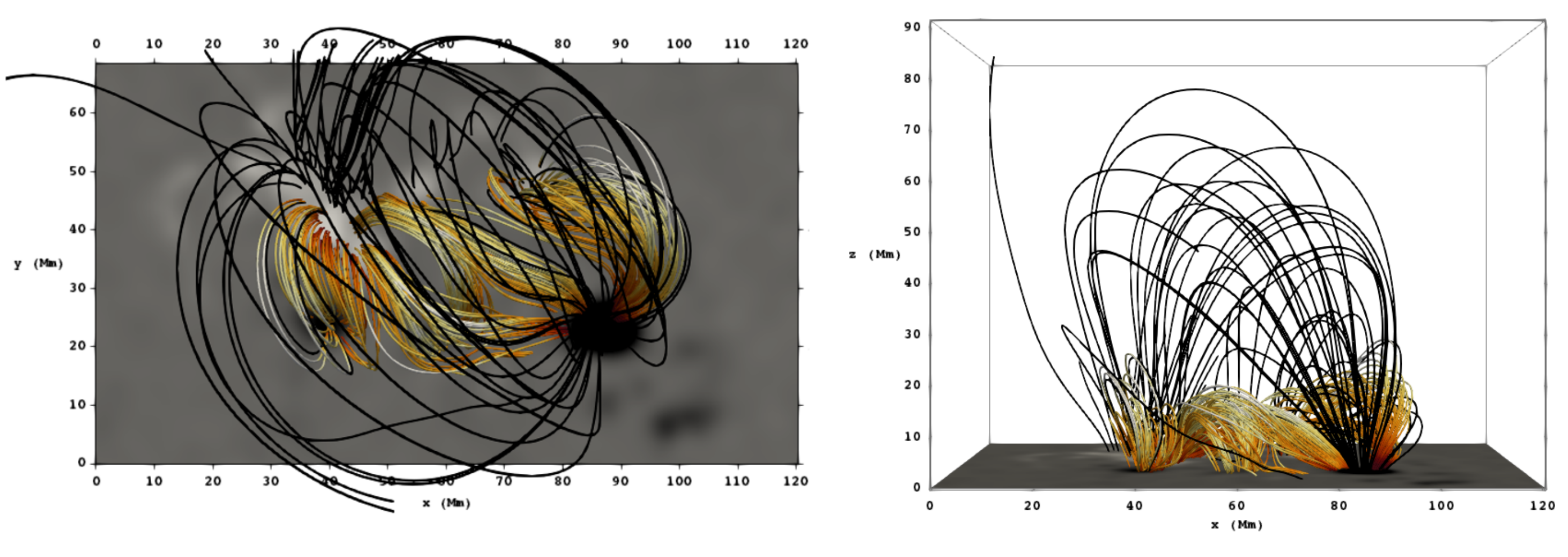}
\put(-275,142){\sf\bf\color{white}(a)}
\put(-180,142){\sf\bf\color{black}(b)}
\end{subfigure}
\begin{subfigure}
\centering
\includegraphics[width=1.0\textwidth]{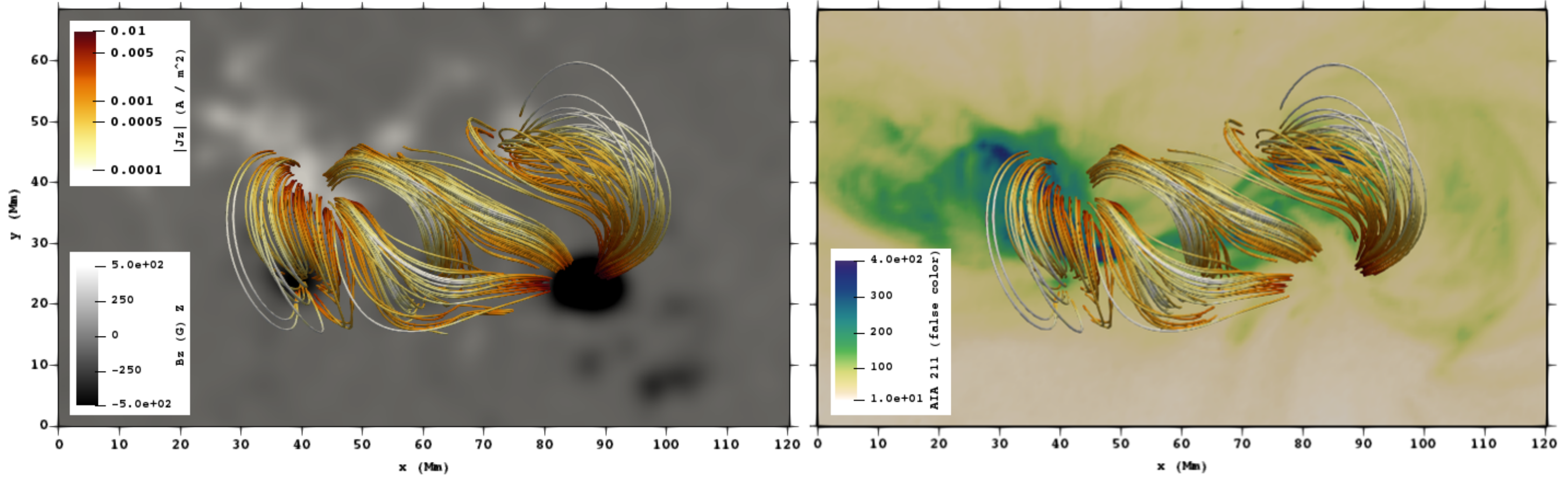}
\put(-275,142){\sf\bf\color{white}(c)}
\put(-245,142){\sf\bf\color{black}(d)}
\end{subfigure}
\caption{NLFF coronal magnetic field of AR~12734 on 2019~March~8 at 02:59~UT (prior to the flare). (a) In addition to the field line connectivity in the AR core (colored), model field lines outlining the overlying field are shown in black. (b) Same as in (a) but when viewed along the positive y-direction. In the bottom panels NLFF coronal magnetic field is displayed on top of (c) the NLFF lower boundary $\Bz$, and (d) a false-color AIA 211~\AA\ image. Model field lines are drawn from randomly selected footpoints and color-coded according to the mean absolute vertical current density at both footpoints.}
\label{fig4}
\end{figure*}

\begin{figure}
\centering
\includegraphics[width=0.45\textwidth]{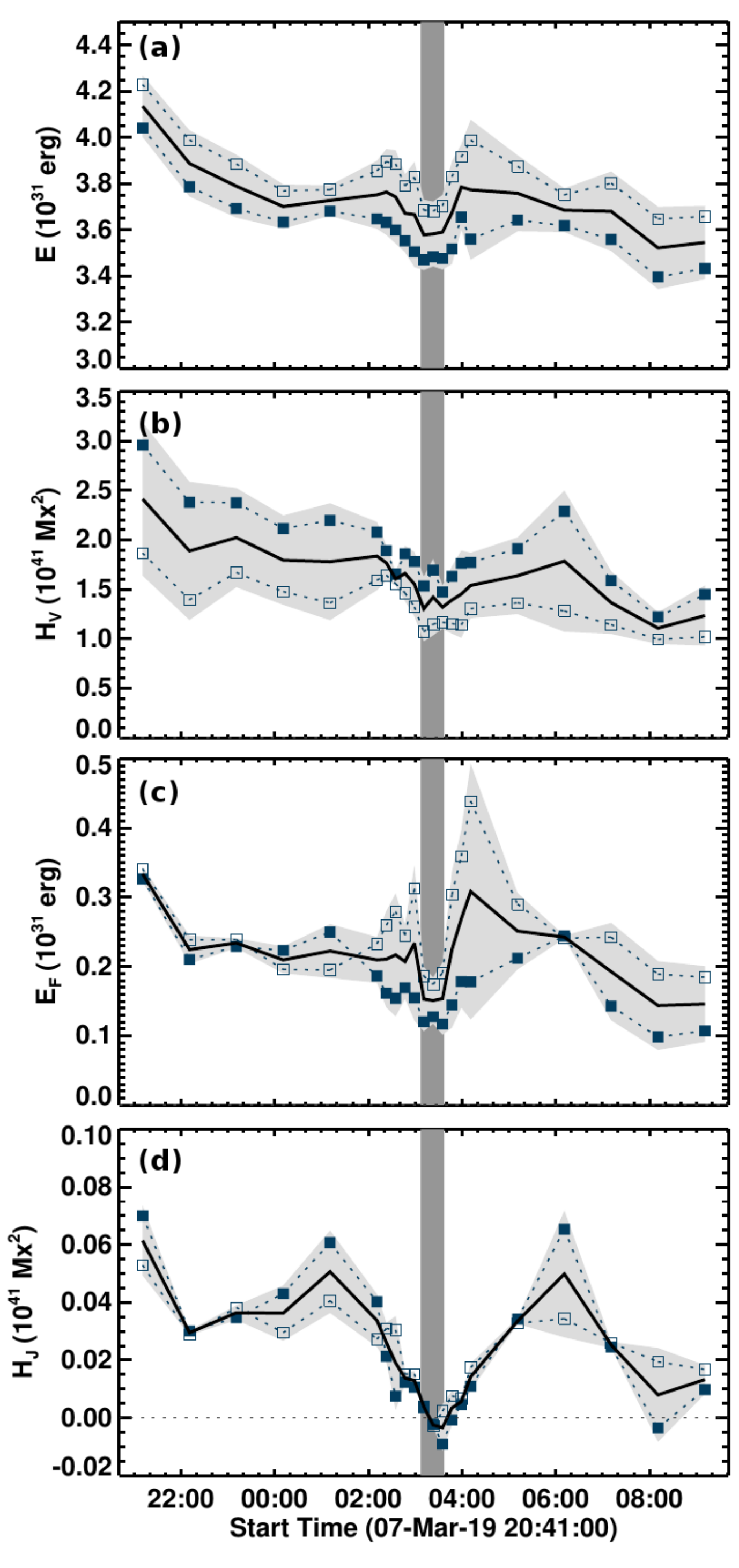}
\caption{Magnetic energies and helicities of AR~12734, deduced from NLFF modeling. Volume-integrated (a) total magnetic energy, $\Etot$, (b) total magnetic helicity, $\Hv$, (c) free magnetic energy, $\Ef$, and (d) magnetic helicity of the current-carrying field, $\Hj$. Different plotting symbols mark time series of NLFF models based on different free model parameter choices. The black solid line represents mean values, and the gray-shaded area indicates the corresponding $1\sigma$-uncertainties (standard deviation). The time cadence of the NLFF model is 12~minutes around the C1.3 flare, and one hour otherwise. Gray-shaded vertical band marks the extended impulsive phase of the flare, as in Fig.\,\ref{fig1}.}
\label{fig5}
\end{figure}

\begin{figure*}
\centering
\includegraphics[width=1.0\textwidth]{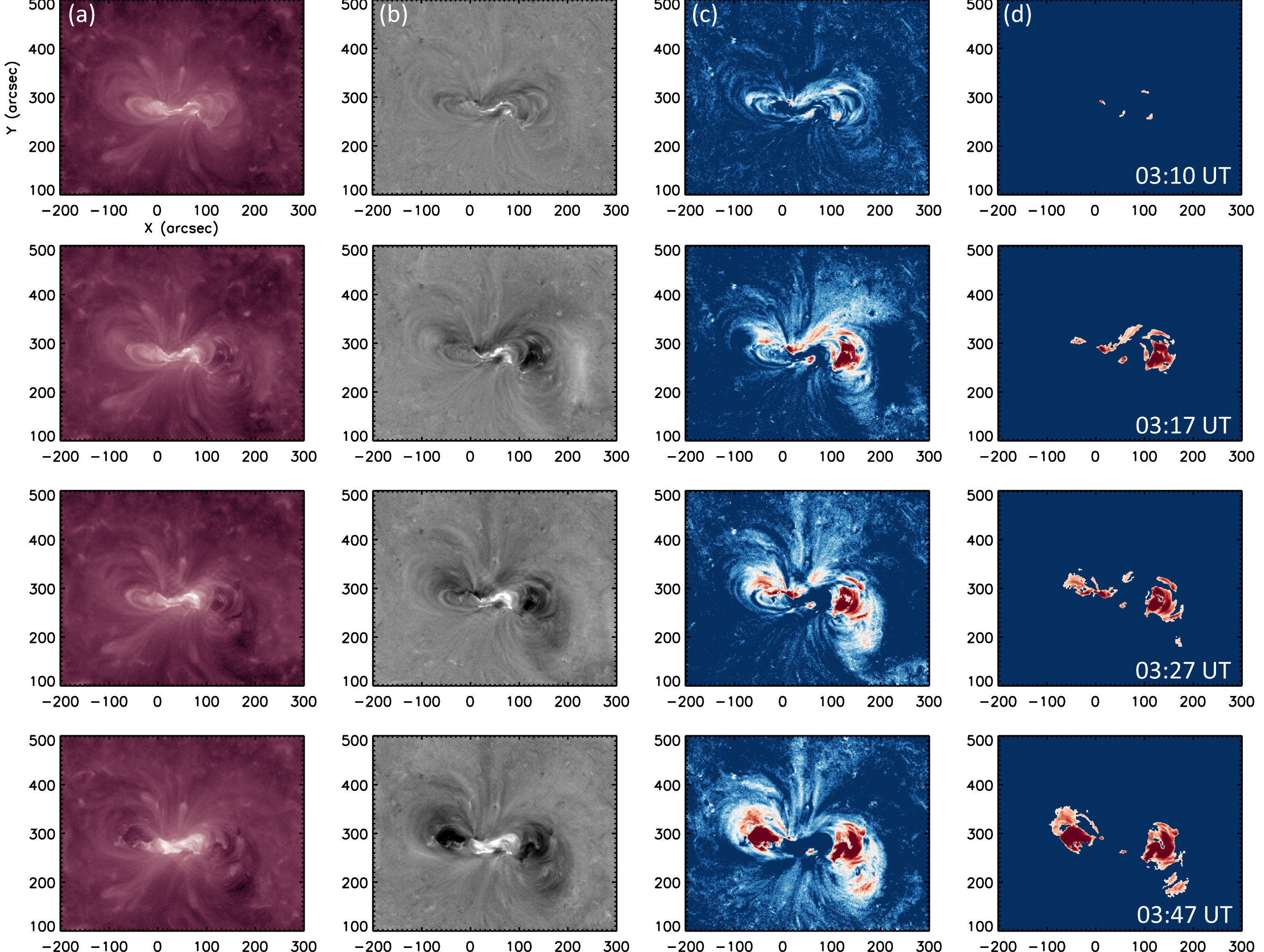}
\caption{Coronal dimming evolution as observed in SDO/AIA 211 {\AA} filtergrams: a) direct images, b) logarithmic base ratio images scaled to $[-1,1]$, c) logarithmic base ratio images scaled to $[-0.7,0]$ to highlight the pixels that decreased in intensity, d) segmented dimming masks applying a  logarithmic threshold level of $-0.38$. A movie accompanying the figure is available in the electronic material (movie3).}
\label{fig6}
\end{figure*}

\begin{figure}
\begin{subfigure}
\centering
\includegraphics[width=0.45\textwidth]{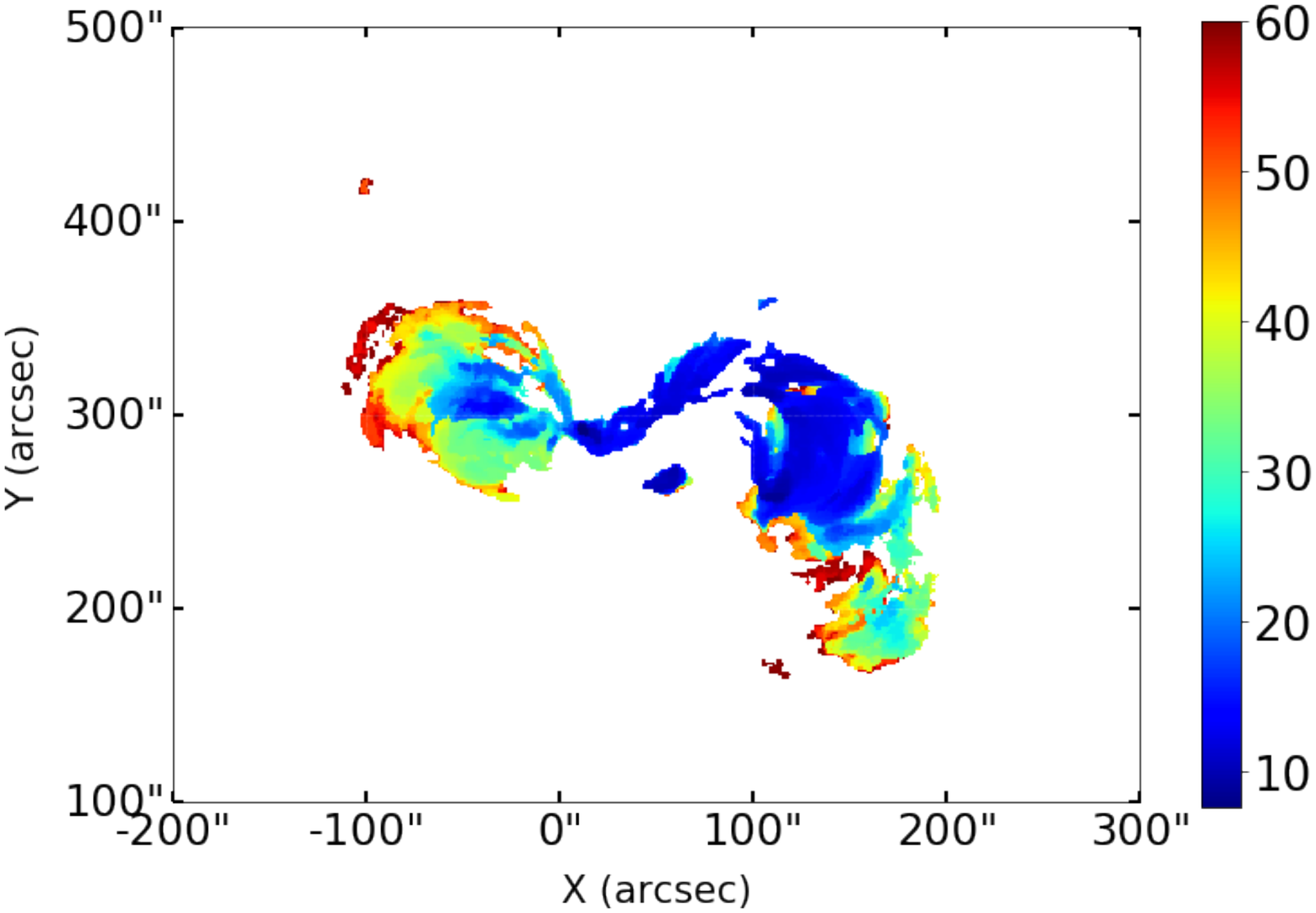}
\end{subfigure}
\begin{subfigure}
\centering
\includegraphics[width=0.5\textwidth]{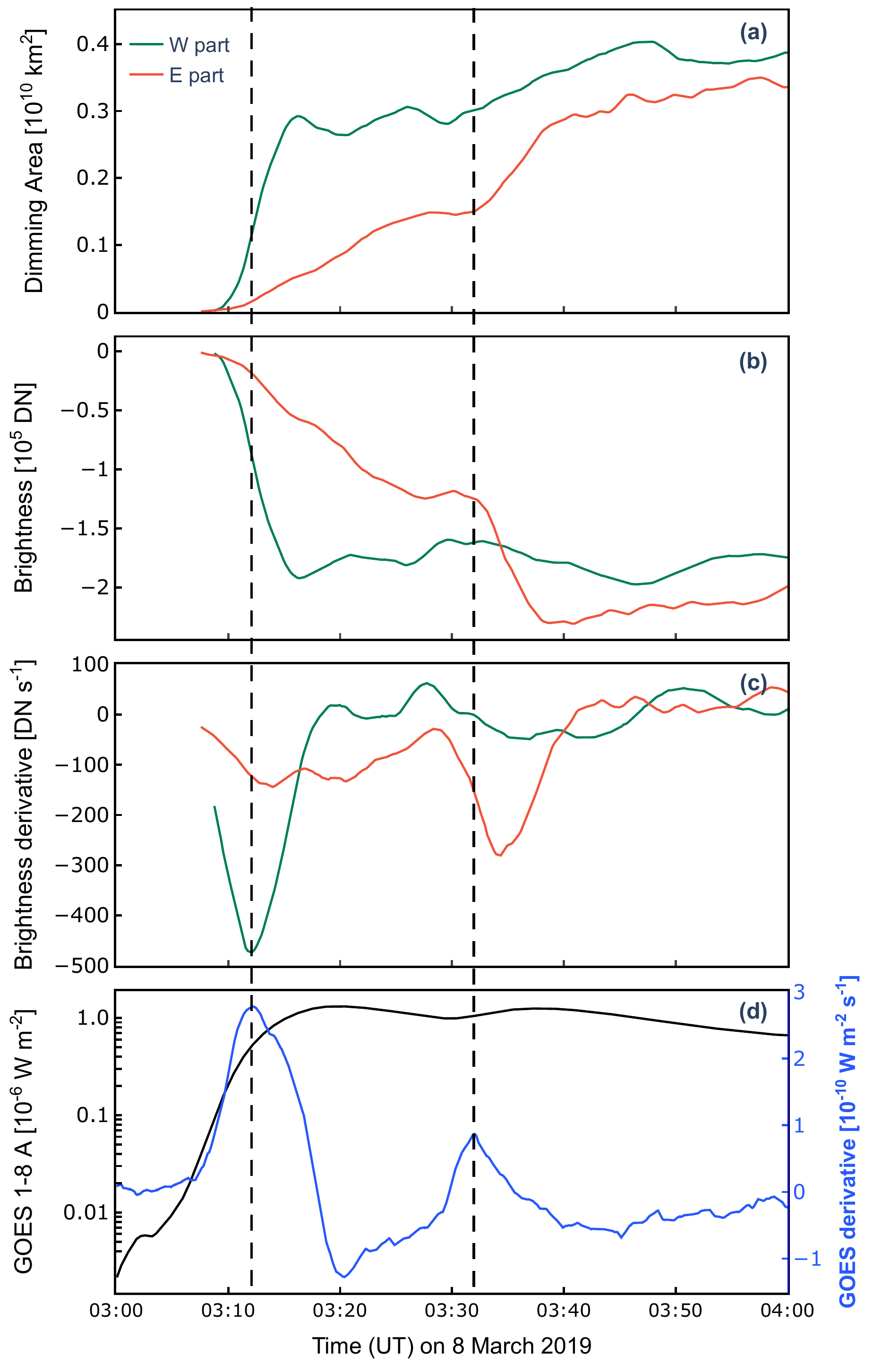}
\end{subfigure}
\caption{\textit{Top panel}: timing map of the coronal dimming region indicating when each dimming pixel was detected for the first time. The color bar in the timing map gives the time (in minutes) after 03:00 UT.
\textit{Bottom panels (a-d)}: time evolution of the derived coronal dimming parameters: a) instantaneous dimming area and b)  brightness,  calculated separately for the  eastern (orange) and western (green) components of the dimming region. c) Time derivative of the dimming brightness plotted in panel b. d) GOES 1--8 {\AA} soft X-ray light curve and 0.5--4 {\AA} time derivative of the  flare emission. Two vertical dashed lines mark the peaks in GOES derivative. The curves are smoothed with a forward-backward exponential smoothing method \citep{Brown1963}.}
\label{fig7}
\end{figure}

\begin{figure*}  
\centering
\includegraphics[width=1.0\textwidth]{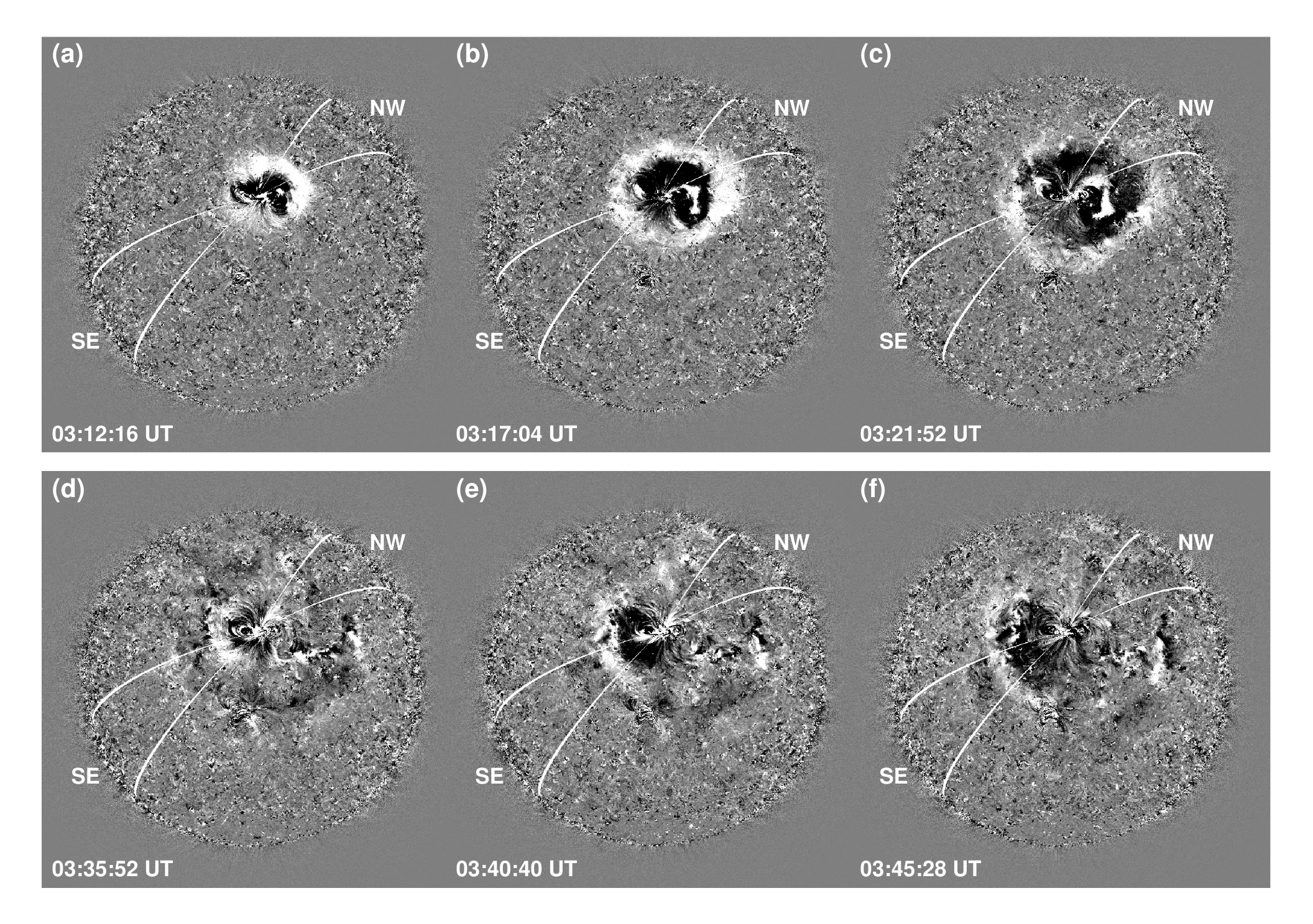}
\caption{EUV wave overview as observed in SDO/AIA 193 \AA~running difference images. Top panels a)--c): Snapshots of the first wave with the quasi-symmetric and quasi-circular propagation around the eruptive center. Bottom panels d)--f): Second wave exhibits asymmetric character of expansion toward SE. We follow the EUV wave in the SE sector where two waves are observed and in the NW sector, where we see only the first wave. A movie accompanying the figure is available in the electronic material (movie4).}
\label{fig8}
\end{figure*}

\begin{figure}
\centering
\includegraphics[width=0.48\textwidth]{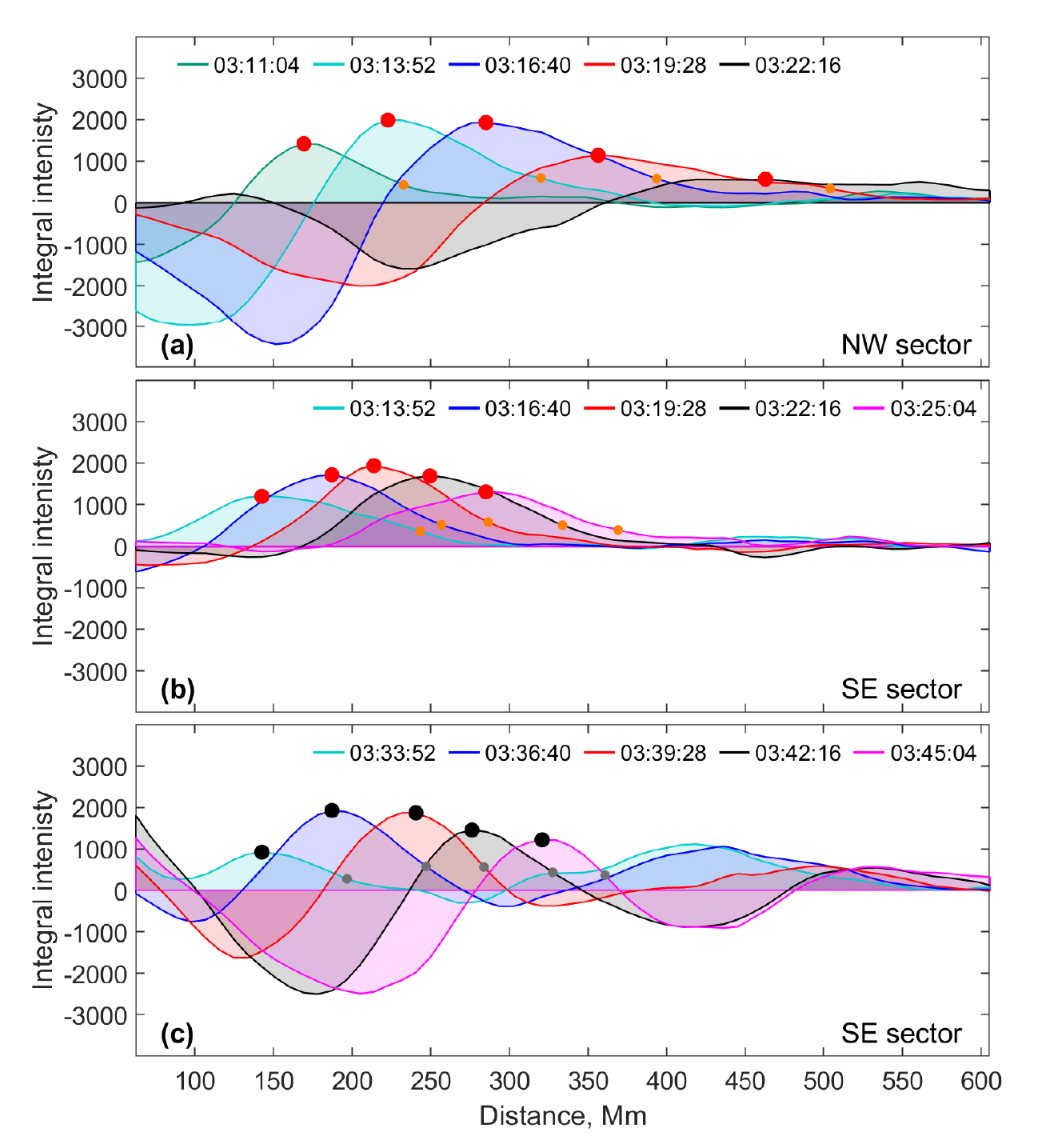}
\caption{Intensity perturbation profiles illustrating the evolution and propagation of the EUV waves in different directions (see sectors plotted in Fig.\,\ref{fig8}): a) first wave in NW sector, b) first wave in SE sector, c) second wave in SE sector. Big circles (red - first and  black - second wave) mark the peak amplitude of each profile, small circles (orange - first and  gray - second wave) mark the determined wavefront positions (extracted at 30\% of the peak value) used for the EUV wave kinematics plotted in Fig.\,\ref{fig10}. A movie accompanying the figure is available in the electronic material (movie5), showing the dynamics of the intensity perturbation profiles with a cadence of 24 seconds.}
\label{fig9}
\end{figure}

\begin{figure}  
\centering
\includegraphics[width=0.48\textwidth]{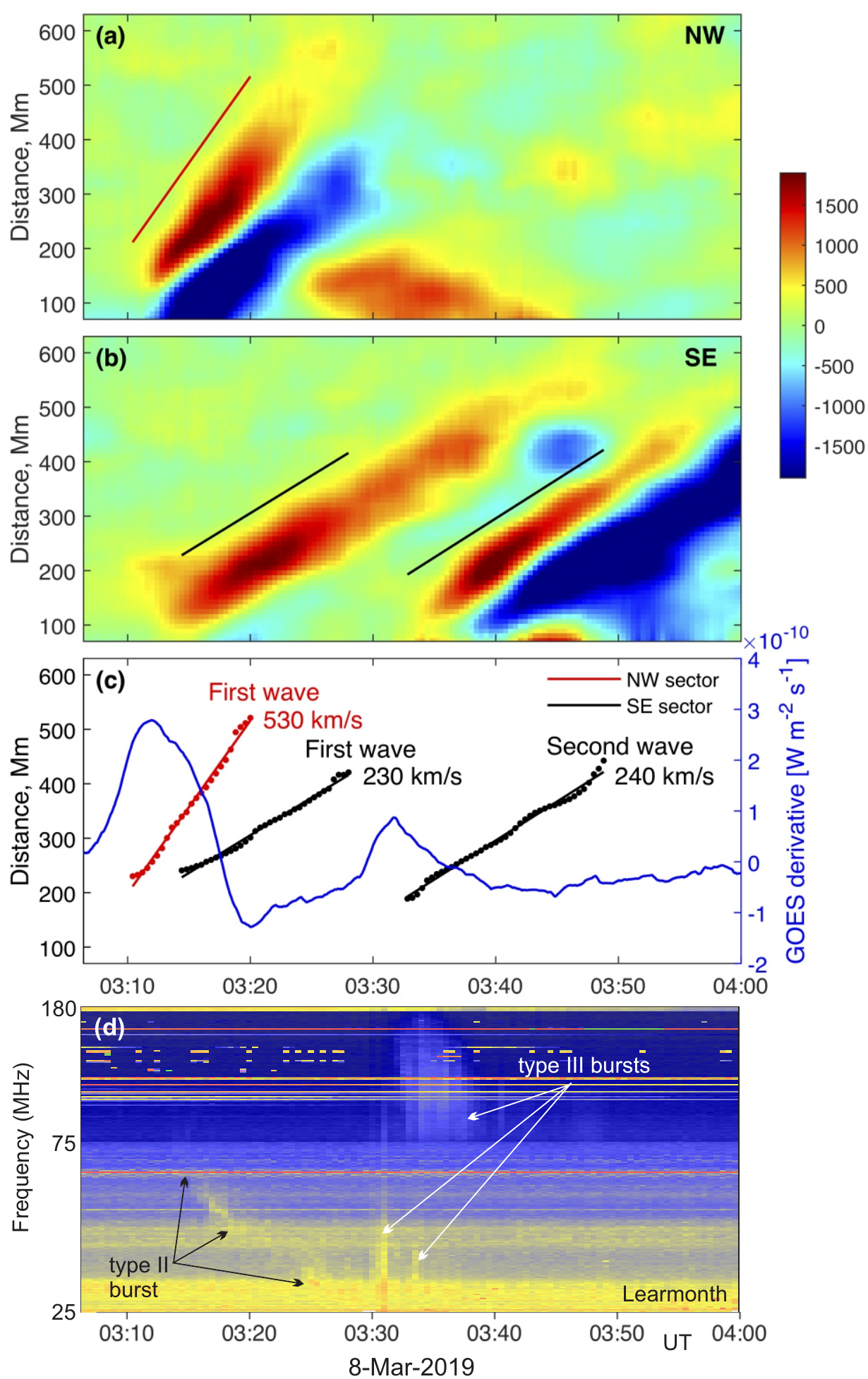}
\caption{EUV wave kinematics. a), b) Stack plots derived from the ring analysis applied to SDO/AIA 193~{\AA} running difference images for the wave segments in the propagation of the NW and SE sector, respectively (indicated in Fig.\,\ref{fig8}). c) Distance-time profiles of the leading fronts of the first EUV wave in the NW (red) and two consecutive EUV waves in the SE sector (black). Dots denote data points, the dashed lines the linear fit. The blue line shows the time derivative of the GOES 0.5--4~{\AA} soft X-ray flux of the associated flare. d) The dynamic radio spectrum recorded by the Learmonth Radio Spectrograph shows the radio emission associated with the EUV waves. The type II radio burst seems to be temporally associated with the first peak and the type III bursts are associated with the second peak of the derivative of GOES soft X-ray flux. }
\label{fig10}
\end{figure}

\begin{figure*}
\centerline{\includegraphics[width=1.0\textwidth]{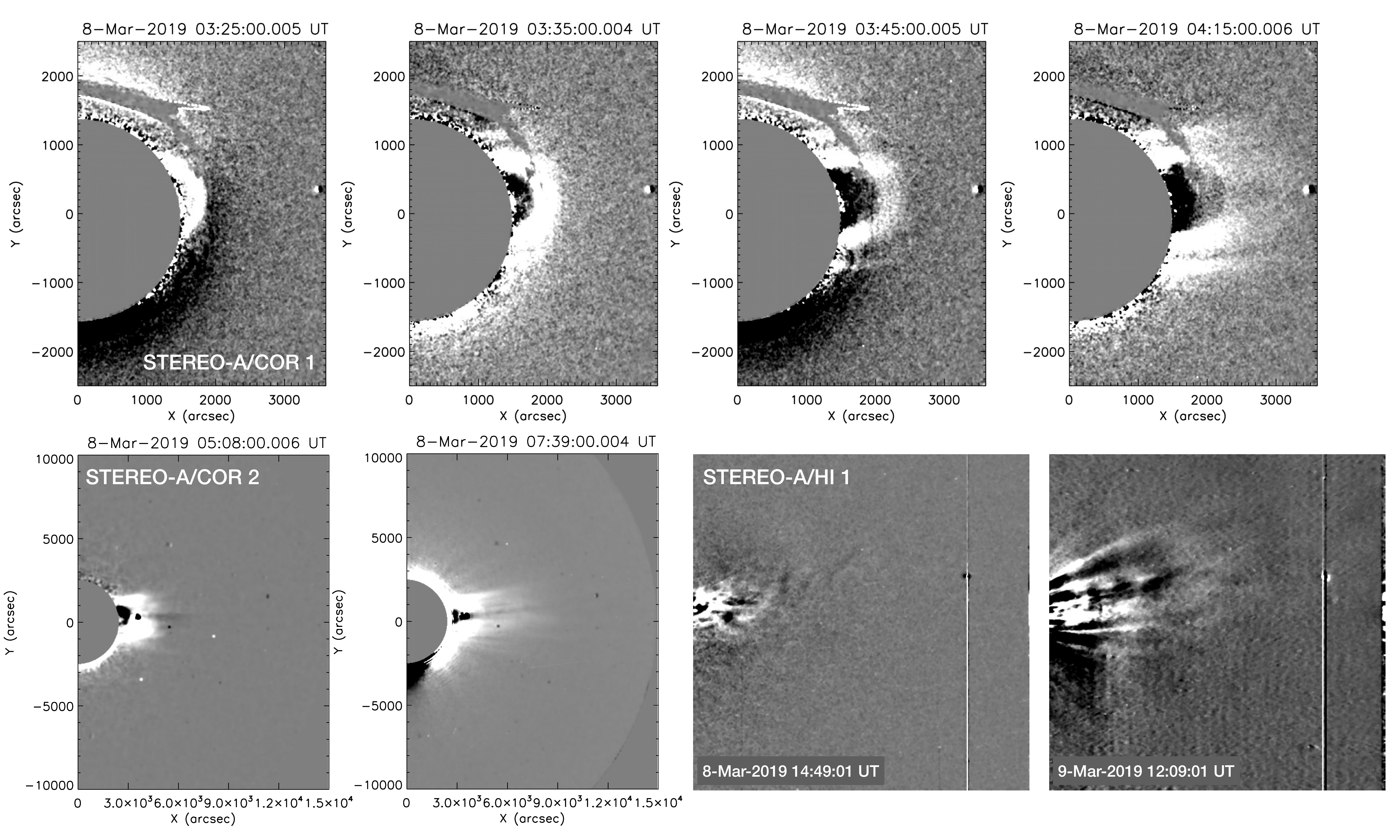}}
\caption{Overview of the CME as seen from STEREO-A in the white-light coronagraphs COR1 and COR2 as well as the heliospheric imager H1.}
\label{fig11}
\end{figure*}

\begin{figure*}
\centering
\includegraphics[width=1.0\textwidth]{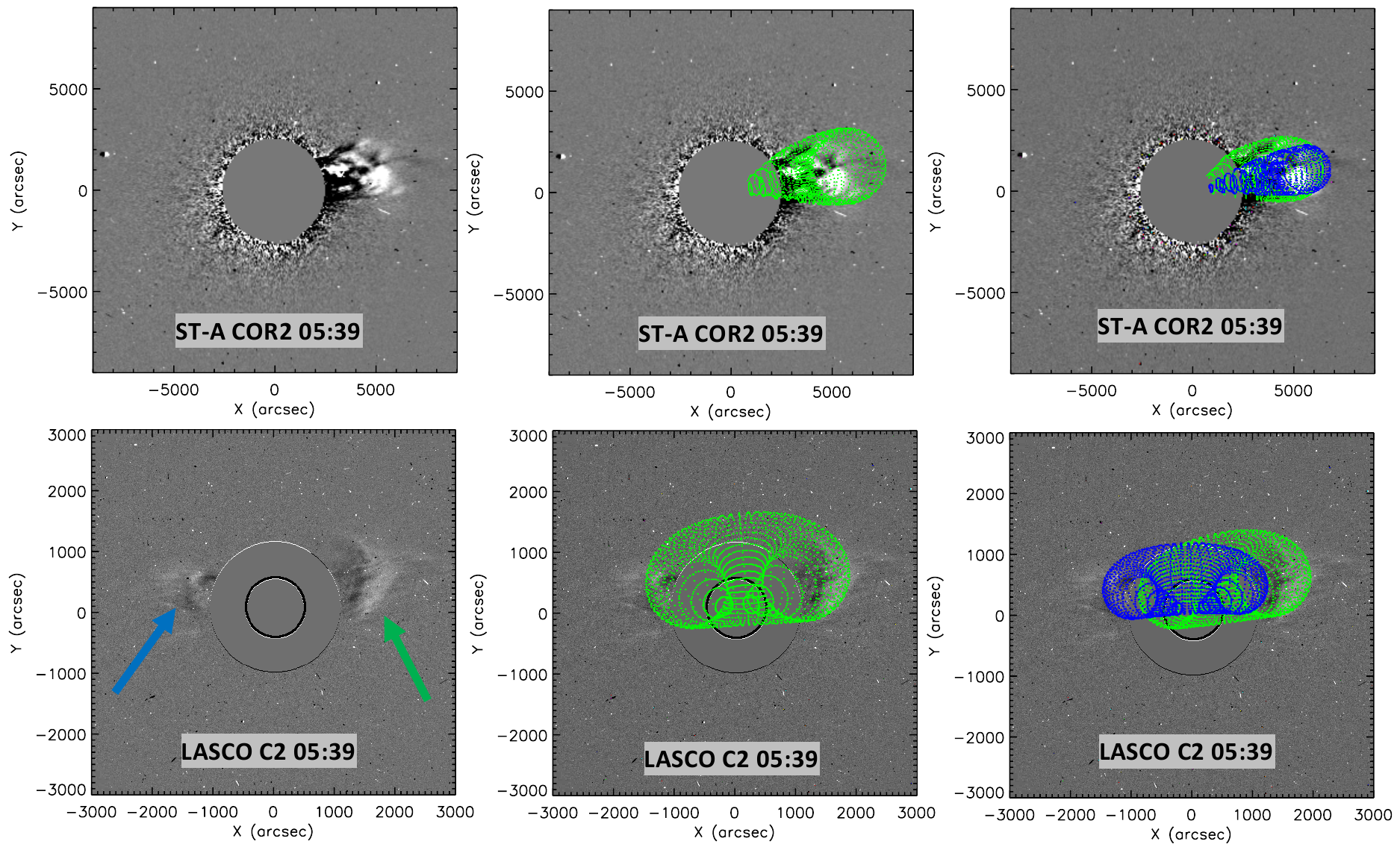}
\caption{The GCS reconstruction of the CME using STEREO-A COR2 and LASCO C2 observations. The blue and green arrows point towards two CMEs listed in the SOHO/LASCO CME catalog (i.e.\, two observed features). The green mesh in the middle panels corresponds to the first GCS reconstruction, assuming a single structure. The blue and green mesh in the right panels correspond to the second reconstruction, assuming two structures. For further details see main text.}
\label{fig12}
\end{figure*}

\begin{figure}
\centering
\includegraphics[width=0.48\textwidth]{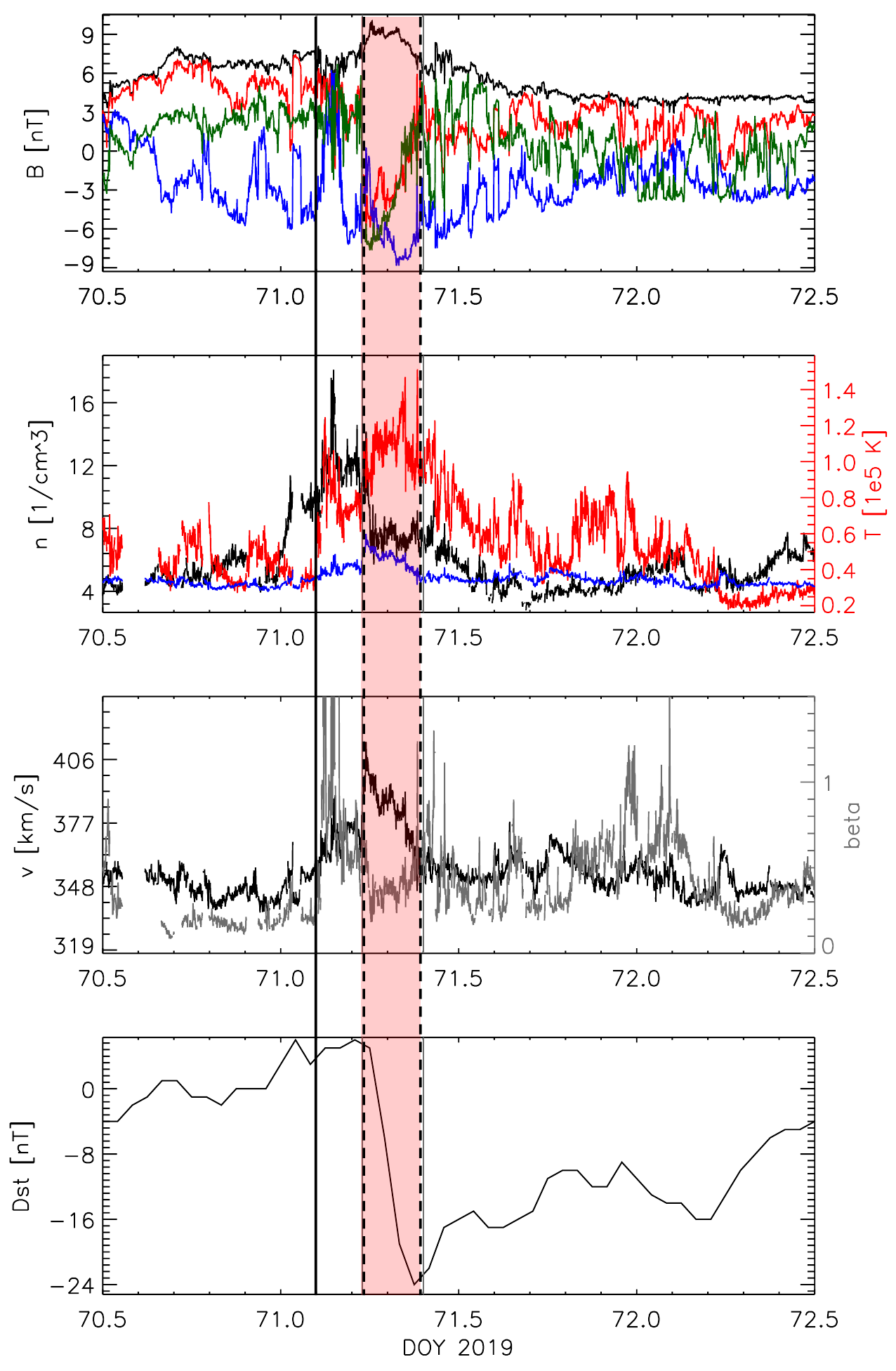}
\caption{In situ measurements given in day-of-year (doy) time series around 2019~March~12 (doy 71). Top-to-bottom panels show: 1) magnetic field strength (black) and its components x, y, z in Geocentric solar ecliptic (GSE) colored red, blue and green, respectively; 2) plasma density (black), temperature (red) and expected temperature (blue); 3) plasma flow speed (black) and beta (gray); 4) Dst index. The expected temperature was calculated according to \citet{lopez87} and \citet{richardson95}. The SFR is highlighted red, with dashed lines outlining its borders. The solid line marks the start of what appears to be a pile-up region ahead of the SFR (for details see main text).}
\label{fig13}
\end{figure}

\section{Observations and modeling related to various aspects of the event}
\subsection{Active Region and flare overview} 
\label{AR}

NOAA AR~12734 emerged during 2019~March~3--4 around the heliographic position N10E40 and during its passage over the solar disk had low flare productivity. During its lifetime, it produced only two small flares: a C1.3 flare on 2019~March~8 (SOL-2019-03-08T03:07C1.3) and a B6.2 flare on 2019~March~9 (SOL-2019-03-09T12:26B6.2). Here we study in detail the C1.3 flare that occurred at N09W13, and the plethora of eruptive phenomena associated despite its small GOES importance class. The flare showed two peaks in the GOES soft X-ray lightcurves (see Fig.\,\ref{fig1}), indicative of two episodes of energy release.

Fig.\,\ref{fig1} shows the time evolution of the soft X-ray (SXR) and ultraviolet (UV) emission of the flare under study. The full-disk GOES 1--8~\AA\ SXR flux (black curve) reveals two peaks, one around 03:18 UT and a subsequent one around 03:37~UT. The double-peak structure is also observed in the area-integrated UV emission derived from SDO/AIA 1600~\AA\ filtergrams (green line). The AIA 1600~\AA\ light curve evolves co-temporally with the time derivative of the GOES 0.5--4~\AA\ SXR flux, indicative of the Neupert effect relating the non-thermal and thermal flare emissions \citep{Neupert1968,Veronig2005,Qiu2021}. The two peaks in these curves are indicative of two episodes in the flare energy release process, separated by about 20~min.

The overall magnetic pattern of AR~12734 can be characterized as quadrupolar, as two bipolar groups can be identified, a leading and a following one (see top panels in Fig.\,\ref{fig2}). This is a relatively rare configuration, as was shown by \citet{2017ApJ...834...56T}. The total area covered by sunspots, at the time when AR~12734 hosted the C1.3 flare, was about 20~MSH, i.e.\,, characteristic for ARs that typically produce only flares below GOES class M1.0 \citep[e.g., Fig.~4 of][]{2017ApJ...834...56T}. The AR contained a total unsigned magnetic flux of $\approx4.5\times10^{21}$~\mx, which is rather small, even for an AR producing only C-class flares and smaller \citep[e.g., Fig.~8 of][]{2017ApJ...845...49K}.

On the day before the C1.3 flare, on March 07 at 07:44~UT, a small filament is observed in synoptic H$\alpha$ images\footnote{\url{http://cesar.kso.ac.at/halpha3s/JPEG/synoptic/2019/kanz_halph_fd_20190307_074448.jpg}} from the Kanzelh\''ohe Solar Observatory. These H$\alpha$ observations support that a single extended (or a number of smaller individual) flux rope(s) existed prior to the onset of the C-class flare, aligned with the solar east-west direction and connecting the leading sunspot in the west of AR~12734 with the extended plage region to its east.

Inspection of the (E)UV emission features within the host AR allows us to establish the link to the underlying flare process (cf. Fig.\,\ref{fig2} and accompanying movie). During stage one of the C1.3 flare, covering the time span $\sim$03:07~UT--03:20~UT, the coronal field of the leading (western) sunspot of AR~12734 was destabilized. Distinct flare kernels appeared first to the north-east (``K1'') and north (``K2'') of the leading sunspot. The north-eastern one developed into an eastward progressing flare ribbon (``R1''). Simultaneously, EUV observations reveal loop-like structures erupting towards solar west (marked by large white arrow in the bottom let panel). During this first stage of the flare, which largely destabilized the western part of the AR, we also notice a narrow ejection that originated from the north-east of the AR, at a location where the observed ribbon R1 roughly terminates (indicated by a small arrow in the AIA 94~\AA\ filtergram at 03:11:36~UT).

However, only during stage two of the flare (after $\sim$03:20~UT), the corona in and around the eastern part of the AR fully destabilized (involving also that part of the AR in which the narrow ejection has been observed before; see right column in Fig.\,\ref{fig2}). During this stage, the flare kernel and ribbon emission that developed during the early stages of the flare further progressed towards solar east, while another flare ribbon formed to the south of it (``R2''). Simultaneously, expanding loop-like structures are observed in the Eastern part of the AR (see large white arrow in the bottom right panel).

These observations indicate that the two successive reconnection events during the double-peak C-class flare are causally connected (see movie accompanying Fig.\,\ref{fig2}), i.e.\,, are indicative of so-called sympathetic flares/eruptions. After the two-stage flare process, during which first the western and subsequently the eastern active-region corona was destabilized, the post-flare corona appears to be still highly non-potential. This is noticeable in the form of apparently highly sheared and/or twisted EUV loops observed after the nominal end time of the flare at 03:58~UT (see movie associated to Fig.\,\ref{fig2}). Importantly, the flare-related reconfiguration appears to have effected mainly the coronal magnetic field that bridged the filament channel, as the latter is still clearly visible (and has even grown) in H$\alpha$ observations\footnote{\url{http://cesar.kso.ac.at/halpha3s/JPEG/synoptic/2019/kanz_halph_fd_20190308_074338.jpg}} at about four hours after the flare occurred. An available online movie, covering the time span $\sim$07:39~UT to $\sim$15:10~UT suggests that the filament channel ``survived'' the eruptions related to the C-class flare under study\footnote{\url{http://cesar.kso.ac.at/halpha3a/2019/20190308/20190308_movie.html}}.

Fig.\,\ref{fig3} shows the evolution of the event as observed by the Extreme-Ultraviolet Imaging Telescope \citep[EUVI;][]{Wuelser2004} onboard STEREO-A \citep{Kaiser:2008}. On 2019~March~8, the longitude of STEREO-A was $-98.1^\circ$ with respect to the Sun-Earth line, and thus STEREO-A observed the eruptions associated with the flare (located at solar N09W13 from Earth view) almost perfectly on the Western limb. The EUVI 195~\AA\ running difference images in Fig.\,\ref{fig3} show two subsequent eruptions, which can be identified as the two loop ejections observed in projection against the disk by SDO/AIA  (see Fig.\,\ref{fig2}). The first eruption observed by EUVI (top panels in Fig.\,\ref{fig3}) relates to the ejection that destabilized the western part of the AR corona, as observed in SDO/AIA, whereas the second one (bottom panels of Fig.\,\ref{fig3}) relates to the ejection that affected mainly the eastern part of the AR, as observed in SDO/AIA. It can be seen in Fig.\,\ref{fig3} that the second eruption initiated higher up in the corona compared to the first one (see also accompanying movie to Fig.\,\ref{fig3}). 

\subsection{AR magnetic structure and evolution} 
\label{NLFF}

We investigate the coronal magnetic field configuration in and around AR~12734 using the static 3D nonlinear force-free (NLFF) model \citep[for a review see, e.g.,][and see Appendix\,\ref{app1} for the method used in this study]{2012LRSP....9....5W}. The pre-flare NLFF model at 02:59~UT reveals different magnetic connectivity domains (Fig.\,\ref{fig4}), which were selected for visualization based on the observations presented in Sect.\,\ref{AR}, i.e.\,, assumed to represent the coronal magnetic field overlying the filament channel observed in H$\alpha$. Thus, the overall shape of the modeled AR core field coincides with the coronal loops observed in the EUV images (Fig.\,\ref{fig2}) which got partially involved during the successive flare-related eruptions.

On the one hand, twisted magnetic field (oriented in a north-southward direction) connects the leading negative-polarity sunspot to the positive-polarity area in the north-west (NW) of the AR (see reddish-colored sample field lines in Fig.\,\ref{fig4}a). On the other hand, a system of sheared field (also oriented in a north-southward manner) connects the north-eastern (NE) plage region to the smaller negative-polarity field in the south-east (SE) of the AR. The two aforementioned systems are connected by magnetic field along the NE-SW direction. The latter appears to represent the main orientation of the underlying magnetic dipole field (the overarching field is represented by the black lines in Fig.\,\ref{fig4}a). In contrast to the much more potential (thus, more vertical) envelope field, the more complex magnetic field in the AR core is concentrated at heights below $\lesssim20$~Mm (see Fig.\,\ref{fig4}b). The magnetic field associated to the leading sunspot (beyond $x\approx70$~Mm in Fig.\,\ref{fig4}c) suggests the existence of a twisted magnetic field of positive handedness, and sheared arcades associated to the plage region covering the eastern part of the AR.

In order to analyze the time evolution of the coronal magnetic field in response to the C1.3 flare, we compute the total and free magnetic energy budgets as well as the relative magnetic helicity, based on a series of NLFF models, during a $\sim12$~hour time interval centered around the C-class flare. The total magnetic energy ($\Etot$) indicates the overall corresponding amount being present in the coronal magnetic field, while the free magnetic energy ($\Ef$) represents an upper limit for the magnetic energy which can be liberated via reconnection processes. The total relative helicity ($\Hv$) is a gauge-invariant measure for the geometrical complexity of the magnetic field within the finite coronal volume \citep[e.g.,][and see Appendix\,\ref{app1} for the computational method used in this study]{1984JFM...147..133B,1985CPPCF...9...111F}. Furthermore, the contribution of the current-carrying (non-potential) magnetic field alone ($\Hj$) can also be meaningfully computed \citep[][]{2003and..book..345B}. In agreement to the visual impression of the core AR magnetic field being sheared and/or positively twisted
prior to the C1.3 flare, we find the total relative helicity, $\Hv$, to be positive in sign (Fig.\,\ref{fig5}b). Similarly, the contribution of $\Hj$ is found to be positive for the pre-flare corona (Fig.\,\ref{fig5}d).

The coronal pre-flare magnetic energy is of the order of $\Etot\simeq3.8\times10^{31}~\erg$ (Fig.\,\ref{fig5}a), and its time evolution reveals a clear response to the C1.3 flare. In particular, we notice a gradual (step-wise) decrease starting already about one hour prior to the start of the flare. After the flare, it takes about one hour to regain a characteristic pre-flare level of $\Etot$. More pronounced changes are found in the time profile of the free magnetic energy, which comprises $\lesssim10$~\% of the total energy (Fig.\,\ref{fig5}c). We observe a sudden decrease of the free magnetic energy during the C1.3 flare by $\Delta\Ef\approx8\times10^{29}~\erg$. 
After the flare, it takes less than one hour to regain, and even to exceed, the characteristic pre-flare level of  $\Ef\simeq2.5\times10^{30}~\erg$, indicating a rather fast replenishment.

The time evolution of the coronal magnetic helicity, $\Hv$, is similar to that of $\Etot$, also showing a step-wise decrease prior to the C1.3 flare (Fig.\,\ref{fig5}b). In contrast to $\Etot$, the replenishment of magnetic helicity appears to happen on a longer time scale, requiring about three hours to regain a characteristic pre-flare level of $\Hv\simeq2.0\times10^{41}~\mxmx$. Notable changes in the time profile of the current-carrying (non-potential) helicity component, $\Hj$, are detected as early as two hours prior to the start of the C1.3 flare, with a value of $\Hj\simeq5\times10^{39}~\mxmx$ about two hours prior to the flare (Fig.\,\ref{fig5}d). Furthermore, we find a flare-related change $\Delta\Hj\approx1.7\times10^{39}~\mxmx$, involving reversal of the sign of $\Hj$. More precisely, we find that the pre-flare corona is characterized by positive values of $\Hj$, which is drastically reduced during the C1.3 flare, resulting in values of $\Hj$ nearly zero just after the extended impulsive phase of the flare. The flare-related changes of $\Hj$ suggest the ejection of magnetic structures that have, on overall, a positive handedness. Analogously as for $\Hv$, also for $\Hj$ it takes about three hours to regain values similar to the pre-flare corona.

For completeness we note that, though not explicitly shown, the NLFF magnetic field models in the post-flare phase own to a complexity of the coronal magnetic field, similar to that of the pre-flare corona. In other words, the post-flare model corona also appears strongly non-potential in nature, with highly sheared and/or twisted magnetic field spanning the active-region core (as observed in coronal observations, described in Sect.\,\ref{AR}).

\subsection{Coronal dimmings} 
\label{dimming}

The coronal dimmings caused by the eruptions associated with the C1.3 flare are analyzed using high-cadence SDO/AIA 211~\AA\ images. To extract the dimming regions, we use the thresholding method applied to logarithmic base-ratio images described in \citet{2018ApJ...855..137D}, dividing each frame of the series by a base image recorded at 02:39 UT. For the segmentation of the dimming masks, we apply a logarithmic  threshold of  $-0.38$, and calculate the corresponding areas on the spherical solar surface. Our focus lies on the spatial distribution of the dimming regions and their time evolution, both  with respect to the two-step behavior of the flare under study.

Fig.\,\ref{fig6} and the accompanying movie show the development of the coronal dimming region in the original SDO/AIA 211 {\AA} filtergrams, the corresponding logarithmic base ratio images and the derived detection masks. The movie shows slow changes already at 03:00 UT. At about 03:10 UT, we can see two distinct dimming kernels low in the corona close to the AR core and two at its Western edge. In particular the Northern kernel at about (x,y)=(20$''$,280$''$) is clearly observed to be the footpoint of erupting large loops (see the AIA 211 \AA\ direct images in the movie). Thereafter, two extended dimming regions are formed, respectively, at the eastern and western edges of the AR, corresponding to the location where the observed flare ribbons terminate (compared to AIA 304 images in third row of Fig.\,\ref{fig2}). Notably, the main dimming evolution also shows a distinct two-step behavior, with first the western dimming region growing and thereafter the eastern one. This characteristics is also clearly seen in the timing map shown in the top panel of Fig.\,\ref{fig7}, where we illustrate the impulsive  evolution phase of the dimming during the first hour of its development. The color coding refers to the time when each pixel was detected as a dimming pixel for the first time.

To analyze the two-step dimming evolution in more detail, we divided the dimming region into the eastern and western parts of the AR (separated at $x=50''$), and tracked the instantaneous dimming area and brightness inside the detected dimming regions separately for both regions (see Fig.\,\ref{fig7}a and Fig.\,\ref{fig7}b). In Fig.\,\ref{fig7}c, we show the time derivative of the dimming brightness. For comparison with the flare evolution, we show the GOES 1--8~\AA\ SXR flux and the 0.5--4~\AA\ flux  time derivative in Fig.\,\ref{fig7}d. It is clearly seen that the evolution of the dimming area and its brightness evolve differently in the western and eastern part of the AR. We see first a fast increase of the dimming area and darkening in the western part, during 03:10 to 03:16 UT, simultaneous with the impulsive phase of the first flare peak in GOES soft X-rays. During this time, the eastern part is also developing but shows a much more gradual evolution. This is followed by an impulsive increase of the eastern dimming from 03:32 to 03:38 UT, in phase with the second peak of the derivative of the GOES flare emission. Therefore, the dimming analysis clearly points to two eruptions. However, we note that we cannot uniquely distinguish core from secondary dimmings in all of the observed dimmings. Thus, from the dimming analysis alone it is not possible to unambiguously conclude whether there are two subsequent eruptions of two different parts of the same structure, or two eruptions of two separate structures.

\subsection{Associated EUV waves}
\label{wave}

Two consecutive EUV waves were observed in close succession associated with the double-peak C1.3 flare under study.  Fig.\,\ref{fig8} and the accompanying movie give an overview of the EUV wave evolution in SDO/AIA 193 \AA~high-cadence (24 s)  running-difference images (created with a lag of 288 seconds). The first wave can be clearly identified after 03:10~UT and shows a quasi-symmetric circular propagation away from the eruptive center (top panels). At 03:12:16~UT (panel a), we already see the prominent formation of the EUV wave in the North-West (NW) direction, while it is not yet so distinct toward the South-East (SE). However, at 03:17:04~UT (panel b) the EUV wave is clearly seen in both the NW and SE directions. At 03:21:52~UT (panel c), the EUV wave is still visible in the SW direction, while there is no longer a clear wave front discernible in the NW direction.
The second EUV wave can be clearly detected for the first time at 03:33:04~UT, i.e.\, around 20 minutes after the initiation of the first EUV wave. As can be seen in the bottom panels of Fig.\,\ref{fig8} (d--f), in contrast to the first one the second EUV wave exhibits an asymmetric character and expands predominantly toward the Eastern direction. The propagation of the two consecutive EUV waves is observed for about 40 minutes.

We study the EUV wave propagation in two selected sectors where the wave(s) show strong signals: (i) in the South-East (SE) sector, where both EUV waves are well observed, and (ii) in the North-West (NW) sector, in which only the first wave is observed. Both sectors have an angular width of 15 degrees. To calculate the kinematics of the EUV waves, we derive intensity perturbation profiles from base difference images using the ring-analysis method \citep{Podladchikova2005,Podladchikova2019,Jebaraj2020}. We first build a spherical polar coordinate system centered at the source AR 12732, and then divide the solar sphere into rings of equal width (8900~km) around the eruptive center. In the next step, we derive the perturbation profiles by calculating the integral intensity of all pixels belonging to each ring in the sector under investigation. The outer border of every ring element is then related to the corresponding distance from the eruptive center. As a result, we derive the projections of the radial intensity profiles onto the surface along the line of sight of SDO.

Fig.\,\ref{fig9} and the accompanying movie show the derived intensity perturbation profiles,  smoothed with a forward-backward exponential smoothing algorithm \citep{Brown1963}, illustrating the evolution and propagation of the EUV waves in the NW and SE directions. The y-axis shows the obtained integral intensity, and the x-axis gives the distance in Mm from the eruptive center. Panels (a) and (b) show the evolution of the first EUV wave in the NW and SE sectors, while panel (c) shows the evolution of the second wave in the SE sector. The EUV wave perturbation profiles are characterized by a sharp increase of the intensity towards its peak followed by a decay to the background level. Big circles (red -- first wave; black -- second wave) mark the peak amplitude of each profile, whereas small circles (orange -- first wave; gray -- second wave) mark the determined wave front positions (extracted at 30\% of the profile's peak value) used for the derivation of the EUV wave kinematics. The high cadence of the SDO/AIA imagery allows us to continuously follow the dynamics of the EUV wave. In the movie, which shows the dynamics of the intensity perturbation profiles with a cadence of 24 seconds, we can clearly see that the first wave propagates much faster in the NW (Fig.\,\ref{fig9}a) than in the SE sector (Fig.\,\ref{fig9}b).

To derive the kinematics of the EUV waves, we extract the positions of the leading wavefronts determined by the 30\% level of the peak amplitude of the  perturbation profiles (indicated by small circles in Fig.\,\ref{fig9}). The resulting kinematics is shown in Fig.\,\ref{fig10}c. Panels (a) and (b) show the stack plots derived from the SDO/AIA 193~{\AA} running difference images and the results of ring analysis for the EUV wave propagation on NW (a) and SE (b) sector. Panel (c) shows the distance-time profiles of the leading fronts of the first EUV wave in the NW (red) and SE (black) sectors, as well as the second EUV wave in the SE sector (black). Dots denote data points, and the dashed lines are the corresponding linear fit. The blue curve in panel (c) shows the time derivative of the GOES 0.5--4~{\AA} soft X-ray flux of the associated flare.

For the first EUV wave in the NW sector, the linear fits gives a mean speed of 530~km/s.  This is more than twice the mean speed derived for the SE sector (230~km/s). The mean speed of the second EUV wave in the SE sector is 240~km/s, which is comparable with the speed of the first wave in the same sector. As can be seen in Fig.\,\ref{fig10}c, the initiation of the EUV waves occurs roughly simultaneously with the peaks of the time derivatives of the GOES SXR emission of the double-peak flare (blue), indicative of the maxima in the flare energy release process \citep{Neupert1968,Dennis1993,Veronig2002}. We also note the spatio-temporal relation of the initiation of the two EUV waves with the two eruptions (ejected loop systems) associated with the two energy release episodes of the flare, indicated by white arrows in the bottom panels of Fig.\,\ref{fig2}: The first (globally) propagating EUV wave is initiated by the first eruption at the western side of the AR, whereas the second EUV wave (propagating only into the eastern direction) is initiated by the second eruption at the East of the AR.

\subsection{Radio signatures and energetic particles}
\label{radio}

The two peaks of the GOES flare co-temporal with the two EUV waves are also associated with two radio events. The dynamic radio spectrum (colour coded time-frequency diagram) observed by the Learmonth Radio Spectrograph shows two radio events observed in close succession (see Fig.\,\ref{fig10}d). Both radio events were simple, consisting of only one type of radio emission.

The first radio event which occurred co-temporally with the first peak of the flare was a faint, but clearly visible type II radio burst indicative of a coronal shock wave \citep[e.g.][]{Wild50,NelsonM85,Vrsnak2008,Mann06,Magdalenic10,Mann18}. Starting at about 03:13 UT, and at approximately 75 MHz, slowly drifting emission lanes of the type II burst are observed in the dynamic spectrum. Type II bursts are considered to be signatures of shock-accelerated electrons generating Langmuir waves at the fundamental electron plasma frequency and/or its harmonics. In this event, the patchy and broken type II band is most probably the fundamental emission. We note the absence of type III radio bursts, i.e.\, radio signatures of fast electron beams propagating along open or quasi-open field lines, which are usually observed before the type II burst and frequently associated with the impulsive phase of the flare \citep[e.g.][]{Magdalenic2014,Magdalenic2020}. The absence of the type III bursts indicates that the event appeared in a region of the corona with mostly closed magnetic field lines.

At about 03:27 UT type III bursts are observed by the Wind/WAVES instrument \citep{1995SSRv...71..231B}, indicating opening of the field lines higher in the corona. This second radio event lasted until about 03:40 UT and consists of only two groups of type III radio bursts. These radio bursts coincide with the second peak of the GOES flare, and they indicate that the configuration of the ambient magnetic field has somewhat changed after the passage of the first EUV wave. The changed topology of the ambient magnetic field provided the possibility for the fast electron beams to propagate along open or large quasi-open field lines and generate observed type III bursts. 

Despite the type III signature of the outward propagating electrons, no electrons or protons have been observed near Earth by the GOES satellite. This is not surprising as the flare/CME eruption is close to the solar disk center while the footpoint of the interplanetary magnetic field of Earth is located towards the west limb of the Sun away from the eruptive AR. In fact, STEREO-A has a better magnetic connection to the AR given its heliospheric location. In situ STEREO-A IMPACT data\footnote{\url{https://stereo-ssc.nascom.nasa.gov/browse/2019/03/08/insitu.shtml}} (not shown here) show an impulsive but small enhancement of 35--65 keV electrons starting shortly before 04:00UT on 2019~March~8, i.e.\,, about half an hour after the initiation of the second EUV wave and the type III bursts. This time delay is in agreement with the timing of relativistic electrons propagating from the Sun to 1 AU following the nominal Parker spiral. No evident enhancement of energetic protons associated with the International Women's Day event has been observed. 

\subsection{CME in the upper corona and interplanetary space}
\label{CME}

The associated CME is analysed in white light observations from the SOHO/LASCO  \citep{brueckner95} C2 and C3 coronagraphs, with a field-of-view reaching 6\,\Rsun\, and 32\,\Rsun, respectively, and STEREO-A(ST-A)/SECCHI \citep{howard08} COR1/COR2/HI1 image data with a field-of-view reaching 2.5\,\Rsun, 15\,\Rsun, and 90\,\Rsun, respectively. Following the two eruptions observed at the west limb in EUVI (see Fig.\,\ref{fig3}), a CME is observed moving into the COR1 and COR2 field of view (FOV), where it is first detected at 03:20 UT and 04:24 UT, respectively. The COR1 and COR2 observations seem to show a single CME. Quickly after appearing in COR2, the leading edge of the CME loses coherence in the radial direction (the latitudinal extent remains relatively well defined) and this feature of a ``diffuse'' leading edge in the radial direction continues in HI1, making it impossible to track the CME leading edge in HI1 via, e.g., J-plots. Nevertheless, HI1 observations indicate a slightly northward direction of the CME. An overview of the CME observations in COR1, COR2 and HI1 is given in Fig.\,\ref{fig11}.

From the Earth perspective,  and including the flare described in Sect.\,\ref{AR}, a CME appears in the C2 FOV at 04:17 UT at a position angle of 300$^\circ$. Quickly after, what appears to be another CME is detected at 04:38 UT at a position angle of 90$^\circ$. These are listed as two different CMEs in the LASCO CME catalog\footnote{\url{https://cdaw.gsfc.nasa.gov/CME_list/}} and marked in the lower left panel of Fig.\,\ref{fig12} by green and blue arrows, respectively. Both features are quite faint in C2 and are no longer reliably traceable in C3 observations. The plane-of-sky linear speeds reported in the LASCO CME catalog for the two features are quite similar, 239 and 294 \kmps, respectively. This indicates that the two structures may correspond to the same CME (especially given the slightly westward location of the source position). Second possibility is that they are two CMEs which already merged and propagate as a single entity.

Further, we perform a 3D reconstruction using the Graduated cylindrical shell (GCS) model by \citet{thernisien06} and STEREO-A COR2 and LASCO C2 images. In GCS model CME is represented by a ``hollow croissant'' geometry, which is fitted manually to COR2 and C2 images to best represent the leading edge of the observed CME. The GCS reconstruction is determined by heliocentric longitude (lon), heliocentric latitude (lat), the height corresponding to the apex of the croissant ($h$), the tilt of the croissant axis to the solar equatorial plane (tilt), the croissant half-angle measured between the apex and the central axis of its leg (half-angle), and the sine of the angle defining the ``thickness'' of the croissant leg (aspect ratio). We first perform GCS fitting under the assumption of a single structure (middle panels of Fig.\,\ref{fig12}), constraining the fit with the location of the AR 12734 as the source region (N09W13). The parameters of this reconstruction are: lon=$5^{\circ}$, lat=$6^{\circ}$, $h$=8\Rsun, tilt=$5^{\circ}$, half-angle=$27^{\circ}$, and aspect ratio=0.3. As can be seen in the middle panels of Fig.\,\ref{fig12}, this GCS reconstruction indicates that the two bright structures observed in LASCO C2 at position angles of 90$^{\circ}$ and 300$^{\circ}$ probably correspond to the same CME. We next perform GCS fitting under the assumption of two structures (right panels of Fig.\,\ref{fig12}), again constraining the fits with the location of the AR 12734 as their source region. The parameters of the fit corresponding to the structure at a position angle of 300$^\circ$ (green mesh in the right panels of Fig.\,\ref{fig12}) are: lon=$15^{\circ}$, lat=$6^{\circ}$, $h$=7\Rsun, tilt=$10^{\circ}$, half-angle=$27^{\circ}$, and ratio=0.27. The parameters of the fit corresponding to the structure at a position angle of 90$^\circ$ (blue mesh in the right panels of Fig.\,\ref{fig12}) are: lon=$-3^{\circ}$, lat=$4^{\circ}$, $h$=7\Rsun, tilt=$3^{\circ}$, half-angle=$27^{\circ}$, and ratio=0.2. As can be seen in the right panels of Fig.\,\ref{fig12}, this GCS reconstruction indicates that the two bright structures observed in LASCO C2 at position angles of 90$^{\circ}$ and 300$^{\circ}$ might correspond to two different structures. However, the overlap indicates that two structures have already merged. Therefore, both reconstructions indicate a single entity evolving further out, regardless of what occurred in the lower corona.

Based on the GCS parameters of the first reconstruction we determine the edge-on and face-on widths of the croissant \citep[see][]{thernisien11} to be $\omega_{\mathrm{min}}=17.5^{\circ}$ and $\omega_{\mathrm{max}}=44.5^{\circ}$, respectively. The corresponding width of the CME in the equatorial plane, estimated taking into account the tilt, using the relation described by \citet{dumbovic19} is $\omega=38^{\circ}$, i.e.\, very similar to the face-on croissant width, as would be expected from the low tilt. Based on the CME longitudinal extent ($\omega=38^{\circ}$) and direction of the CME apex (lon=$5^{\circ}$), we would expect to observe CME signatures near-Earth. However, due to the slightly northward direction of the CME (lat=$6^{\circ}$) and the fact that it is a relatively ``slim'' croissant ($\omega_{\mathrm{min}}=17.5^{\circ}$), we might expect that only the south part of the CME grazes over the Earth. We calculate the radius of the croissant at the apex using equations provided by \citet{thernisien11} and obtain a radius of $1.85$\,\Rsun\, at a distance of 8\,\Rsun. The linear speed of the CME apex determined based on the reconstruction at three different time-steps and assuming self-similar expansion is $v_0=350$\,\kmps. This means, the CME is quite slow, and therefore not likely to pile-up a significant sheath or drive an interplanetary shock \citep[see e.g.,][]{2008SoPh..253..237Z}.

\subsection{In situ observations and Earth impact}
\label{ICME}

The arrival of the 2019~March~8  CME at Earth was indicated by the GEOALERT message from the Solar influences data analysis centre (SIDC)\footnote{\url{http://sidc.be/archive}} on 2019~March~12, as well as the Community coordinated modelling center (CCMC) CME Scoreboard\footnote{\url{https://kauai.ccmc.gsfc.nasa.gov/CMEscoreboard/PreviousPredictions/2019}}. SIDC reports the ICME arrival at 05:44 UT, identified as a speed jump from about 360 to 400 \kmps and the abrupt change of the magnetic field orientation, while CCMC reports ICME arrival at 02:00 UT with unspecified ICME signatures. We note that on 2019~March~12 no interplanetary shock was reported in the Harvard and Smithsonian Center for Astrophysics database (CfA IP Shock Database\footnote{\url{https://lweb.cfa.harvard.edu/shocks/}}), and no ICME was reported by the continuously updated \citet{richardson10} ICME catalog\footnote{\url{http://www.srl.caltech.edu/ACE/ASC/DATA/level3/icmetable2.htm}}.

To investigate the arrival of the interplanetary counterpart of the CME (ICME), we check in\,situ~ measurements of the solar wind plasma and magnetic field parameters, respectively. Fig.\,\ref{fig13} shows data from the Advanced Composition Explorer satellite \citep[ACE;][] {stone98} Solar Wind Electron, Proton, and Alpha Monitor \citep[SWEPAM;][]{mccomas98} and Magnetometer \citep[MAG;][]{smith98}. The data are obtained trough NASA Space Physics Data Facility (SPDF\footnote{\url{https://spdf.gsfc.nasa.gov}}). We supplement these with the time-series of the Disturbance storm time (Dst) index, which is derived from changes of the horizontal component of the geomagnetic field and is provided by the World Data Center (WDC) for Geomagnetism, Kyoto\footnote{\url{http://wdc.kugi.kyoto-u.ac.jp/wdc/Sec3.html}} \citep{WDCkyoto}.

As can be seen in Fig.\,\ref{fig13}, the in\,situ~ measurements around 2019~March~12 (doy 71) are not conclusive of a typical shock-sheath-magnetic cloud ICME arrival. There is an indication of a flux rope structure seen as the rotation in the $B_z$ component of the magnetic field starting at DOY 71.26 and lasting until DOY 71.4 and accompanied by the slightly increased total magnetic field (highlighted red in Fig.\,\ref{fig13}). However, other typical signatures of a magnetic cloud such as low temperature and low plasma beta \citep[e.g.][]{zurbuchen06,kilpua17} are lacking. Given the short duration of this structure ($\approx$ 3.5 hours), such signatures are indicative of a small flux rope \citep[SFR, see e.g.][and references therein]{moldwin00, janvier14a, hu18, murphy20, chen20}. Such magnetic structures in the observed solar wind data are far more frequent than coronagraphic CME observations \citep{janvier14a} and thus debated about whether they are of solar or solar wind origin \citep[i.e.\, formed around the heliospheric current sheet, HCS, ][]{feng08,cartwright10,tian10,janvier14b,murphy20}. We note that around the observed SFR there are no typical signatures of the HCS crossing, such as the reversal of the $B_x$ and $B_y$ GSE components and change in the pitch angle of the suprathermal strahl electrons (not shown here) accompanied by an increased density of the heliospheric plasma sheet \citep{blanco06,sanchez-diaz19}. Moreover, we check Leif Svalgaard's list of the sector boundary crossings\footnote{\url{http://wso.stanford.edu/SB/SB.Svalgaard.html}}, which reports sector boundary crossings on 2019-02-27 and 2019-03-18, i.e.\, nowhere near the observed SFR. Therefore, the observed SFR is likely of solar origin and corresponds to the eruptions associated with solar event SOL-2019-03-08T03:07C1.3).

In the observed SFR the $B_z$ component rotates from $\approx -7.7$ nT up to $\approx 1.6$ nT, i.e.\, we only observe a partial rotation. Nevertheless, it is observed that $B_z$ rotates south-to-north, while the $B_y$ component is negative, indicating counterclockwise rotation in the $B_z$-$B_y$ plane, typical for a right-handed SWN type of flux rope \citep{bothmer98,palmerio18}. This is in agreement with the positive helicity as determined from the NLFFF modeling of the source AR. The size of the SFR, determined based on its duration and estimated mean plasma speed within the structure of 380 \kmps is $\approx 0.03$ AU (6.6 \Rsun). Comparing this to the size of the structure reconstructed using GCS and assuming that the GCS-reconstructed structure expanded self-similarly following a power-law behaviour \citep{bothmer98,demoulin09}, we obtain an expansion power-law index of $n_a=0.18$, which is very low even for SFRs \citep[that typically have $n_a\approx0.4$, see][]{cartwright10,janvier14b}. Therefore, assuming that the observed SFR corresponds to the analysed solar event, it seems that its expansion was severely hampered. Moreover, the plasma temperature and density in and around the SFR seem to be increased compared to the ambient solar wind plasma, and the plasma beta ahead of the SFR is also notably increased. These characteristics indicate strong interactions of the SFR with the ambient plasma.

It can be seen in Fig.\,\ref{fig13} that the plasma speed increase starts before the SFR at $\approx$ DOY 71.05 and the plasma density even earlier, around DOY 71.0. These are followed by a sharp increase of temperature and plasma beta at DOY 71.1 (marked by the vertical solid line in Fig.\,\ref{fig13}). The increased values of plasma density, temperature, speed and plasma beta persist until the frontal border of the SFR, and are accompanied by fluctuating components of the magnetic field, but the total magnetic field is not increased (on the contrary, it even shows signs of local depressions). Therefore, the region in front of the SFR does not have typical sheath/pile-up signatures, possibly also related to strong interactions of the SFR with the ambient plasma ahead.

The last panel in Fig.\,\ref{fig13} shows the Dst index, which is a measure of the geomagnetic response. A clear drop in the Dst index can be observed corresponding to the south oriented $B_z$ component of the interplanetary magnetic field, as would be expected in geomagnetic storms. However, with a minimum value of $-$24 nT, the corresponding Dst index is below the $-$30 nT threshold to be considered even a minor storm \citep[e.g.][]{gonzales94}.

Finally, we note that based on the observed partial rotation of the SFR, a slightly northward direction of the eruptions and GCS-estimated spatial extent it is possible that the structure only grazed over the spacecraft with its south part. Namely, the position of the ACE satellite on 2019~March~12 was $-7.3^\circ$ with respect to the solar equator. Therefore, the latitudinal separation of the CME apex as obtained from the GCS reconstruction to the ACE position is $13.3^\circ$, i.e.\, only $4.2^\circ$ smaller than the edge-on width of the croissant.

\begin{figure}
\centering
\includegraphics[width=0.48\textwidth]{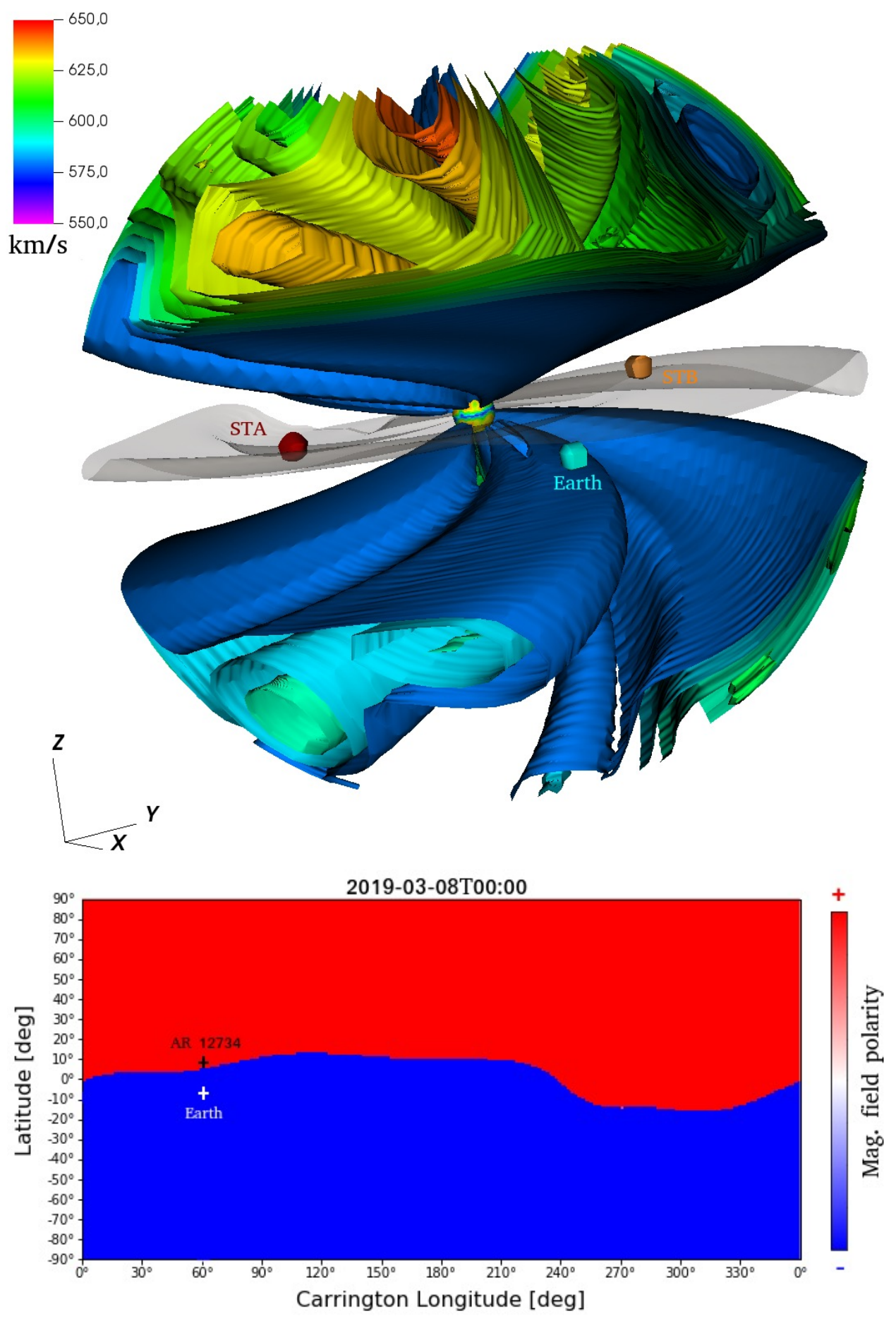}
\caption{The heliospheric current sheet and fast solar wind as modeled by EUHFORIA. The top panel shows modeled fast solar wind and HCS in 3D presentation together with STEREO-A, STEREO-B and Earth. The colorful sphere in the middle of the domain shows the solar wind velocities at the inner boundary of EUHFORIA (at a distance of 0.1 AU) while the colorful isosurfaces depict velocities between 580--648 km/s. The HCS is represented by the gray shaded surface. The bottom panel shows the corresponding magnetic field polarity map, with the HCS at the border of positive (red) and negative (blue) polarity and marked positions of the AR12734 and Earth.}
\label{fig14}
\end{figure}

In order to get a better insight into the state of the heliosphere during the time of the event under study, we perform a simulation using the European Heliospheric Forecasting Information Asset \citep[EUHFORIA,][]{pomoell18}. EUHFORIA is a 3D MHD model consisting of two parts, the coronal (extending up to 0.1 AU) and the heliospheric part. The coronal part of EUHFORIA uses magnetograms as an input to model the solar corona and to provide the inner boundary conditions for the heliospheric part of the model. We employed the GONG Air Force Data Assimilative Photospheric Flux Transport Model (ADAPT) magnetograms as an input and the default EUHFORIA coronal setup \citep{hinterreiter20,samara21}. In this setup the initial MHD solar wind parameters at the inner heliospheric boundary are provided by the semi-empirical Wang-Sheeley-Arge (WSA) model \citep{arge03}, in combination with the potential field source surface (PFSS) model \citep{altschuler69} and the Schatten current sheet (SCS) model \citep{schatten69}. The 3D presentation of the heliospheric current sheet (HCS) and a range of fast solar wind velocities between 580 - 648 km/s as modeled by EUHFORIA, is shown in Fig.\,\ref{fig14}. The relative positions of STEREO-A, STEREO-B spacecraft and Earth are also included. In addition, we show a corresponding magnetic field polarity map, indicating the positions of AR12734 and Earth relative to the HCS. Fig.\,\ref{fig14} shows a very flat HCS passing above Earth, i.e.\, Earth is on the opposite side of the HCS compared to the AR. As discussed in the next section, such configuration can have multiple consequences for CME evolution as well as particle propagation.

\section{Discussion of the results and interpretation of the event}
\label{interpretation}

Both strong and weak flares are related to restructuring of the coronal magnetic field and the sudden release of previously stored free magnetic energy. However, for weak flares this restructuring is usually not so clearly observed. For ARs with major flare activity (GOES class M1.0 and larger), it is generally hard to detect changes of the coronal (magnetic field) evolution in response to C-class flaring. This is because the magnetic field in major flare-productive ARs is generally very complex (Hale classes $\gamma$--$\delta$) and extended (total area covered by sunspots $\gtrsim$200~MSH) \citep[see Table~1 and Fig.~4 of][respectively]{2017ApJ...834...56T}. Then, small flares (GOES class C and smaller) often affect only a minor portion of the coronal volume of the host AR and, e.g., flare-associated changes are nearly indistinguishable with respect to the total (by orders of magnitude larger) magnetic energy and helicity budgets. In the present case, however, we deal with a small AR of moderate magnetic complexity (Hale class $\beta$, total area covered by sunspots $\approx$20~MSH). This allowed us to unambiguously identify the coronal volume involved in the flare-related processes and to quantify the corresponding changes in time profiles of the coronal magnetic energy and relative helicity, deduced from NLFF modeling. Furthermore, its (magnetically) isolated position on the solar disk also allowed us to study a manifold of accompanying eruptive features (EUV waves, coronal dimmings, CMEs, radio bursts, etc.) in great detail.

A flare-related eruption of a pre-existing flux rope may appear in different flavors \citep[for a review see][]{2020RAA....20..165L}. It may erupt as a whole, or only partly \citep[for observation-based studies see, e.g.,][respectively]{2012ApJ...748...77S,2015ApJ...812L..19L}, and even different portions of a flux rope may be ejected during subsequent eruptions \citep[for observation-based studies see, e.g.,][]{2013ApJ...769L..25C,2017SoPh..292...93T}. Destabilization of a certain portion of the AR can modify the overlying field and change the stability of neighbouring magnetic configurations, thus allowing for subsequent eruptions. Indeed, such a scenario has been shown in MHD simulations to account for magnetically coupled sympathetic eruptions \citep[][]{Torok2011}. We interpret the observations of the International Womens Day event by such sympathetic-eruption scenario. First, the two-step nature of the flare energy release profile is evidenced by the SXR flux derivative and the AR-integrated UV emission (Fig.\,\ref{fig1}), as well as the subsequent loop eruptions observed in SDO/AIA EUV imagery (Fig.\,\ref{fig2} and accompanying movie) on the East and West side of the AR and the related two-step coronal dimmings. Second, the fact that the individually involved portions of the coronal magnetic field were commonly located underneath a single large-scale (envelope) field (Fig.\,\ref{fig4}a,b), supports the idea that the stabilizing effect of the overlying field changed in the course of the two-step flare process. The recent stability analysis of \citet{joshi21} suggests that the confinement of the overlying magnetic field prior to the flare onset  was weakest above the western part of the AR, i.e.\,, cospatial to the apparent location of the first eruption, prone to torus instability. We assume that the changes to the confining field due to the first eruption gave way to subsequent setting in of torus instability in the eastern part of the AR, i.e.\,, at the location where the second eruption occurred.

In the present case, we argue that it is the low-corona magnetic field (which bridges a possibly existing low-lying magnetic flux rope) along the main polarity inversion line (PIL) which got successively removed in the two-step flare process, rather than the low-lying flux rope itself. This is supported by the H$\alpha$ observations which show that the PIL-aligned filament is still present after the two-step eruption (see Sect.\,\ref{AR}). This interpretation contrasts that of, e.g., \citet{joshi21}, of a zipping-like asymmetric eruption of a single underlying flux rope, also because such a single pre-existing flux rope is not identifiable from our NLFF models, and the corresponding reconstructed 3D CME geometry is inconclusive in that respect, as discussed at the end of this section.

The two-step flare process had severe impacts on the coronal relative magnetic helicity budget. In particular, the positive, non-potential (current-carrying) helicity dropped to nearly zero in the course of the flare, supporting the ejection of highly non-potential magnetic structures of positive handedness (Fig.\,\ref{fig5}d). Simultaneously, the free magnetic energy dropped by a factor of $\sim$two (Fig.\,\ref{fig5}c), i.e.\,, about $8\times10^{29}$~erg were available to be released. This value is in good agreement with the energy typically released during C1-class flares (about $6\times10^{29}$ erg; e.g., \cite{kretzschmar2011} and Fig.~13 of \cite{tschernitz2018}). 

In association with the two-step flare process, also two EUV waves were observed that occurred in close succession. They were initiated close in time to the two peaks of the GOES SXR flux derivative of the C1.3 flare and two distinct loop eruptions  observed low in the corona in SDO/AIA EUV filtergrams. While the first wave propagated globally in all directions, as has been observed before in solar minimum events \cite[e.g.,][]{thompson1998,Veronig2008,Kienreich2009,Podladchikova2019}, the second one was limited to the East side of the source AR.

The global propagation of the first EUV wave can be explained by the asymmetric eruption on both the East and West side of the AR, with the major part (ejected loop system) into the West and a weaker, narrow part to the East. This asymmetry in the eruption may also explain the very different EUV wave speeds observed for different propagation directions of the first wave (530 km/s in the NW sector; 240 km/s in the SE sector), similar to the asymmetric event studied in \cite{2011SoPh..273..421T}. Many studies have demonstrated that EUV waves are large-amplitude, sometimes even shocked, fast magnetosonic waves propagating through the corona \citep[e.g.,][]{Mann1999,Warmuth2001,Patsourakos2009,Veronig2010,Veronig2018,long17a}. In this scenario, a stronger driver (eruption) initiates a fast-mode MHD wave with a larger amplitude, and due to the non-linear evolution of the large-amplitude  perturbation, a higher amplitude is related to a higher phase speed \citep{Mann1995,Lulic2013,Warmuth2015,Vrsnak2016}.

The second EUV wave was initiated by the ejected loop system at the East edge of the AR and was able to propagate freely in this direction, whereas toward West its propagation was impeded by the high Alfv\'en speeds in the core of the AR. Finally, we note that the first EUV wave was associated with a type II radio burst, indicative of shock formation somewhat higher in the solar corona \citep[e.g.,][]{Mann1999,Vrsnak2008,Warmuth2015}, whereas the second event was associated with type III bursts indicative of only accelerated electrons having access to``open'' field lines, and no associated shock wave.

Although the low coronal signatures clearly indicate two distinct eruptions, white light observations do not unambiguously indicate two separate CMEs. While the SOHO/LASCO view does reveal two segments (in East and West), STEREO-A white light observations do not give any indication of two separate CMEs. A 3D reconstruction using single CME geometry indicates that the two segments observed in LASCO C2 may be flanks of the same CME, whereas a 3D reconstruction using two separate CME geometries indicates their merging. Therefore, we believe that the white light observations point to a single entity, most likely composed of two merged structures. We note that the white light signatures are quite weak, especially in the radial direction around the CME apex as observed by STEREO-A (the leading edge quickly dissolves). The observed structures were launched from a centrally located AR and accompanied by a shock (type II radio burst). Thus, the question arises, why are the white-light signatures so weak, even when viewed with an almost perfect quadrature perspective from STEREO-A? During solar minimum phases, the HCS is known to reside at the low-latitude streamer belt, exhibiting almost no warps \citep[e.g.,][]{2001JGR...10615819S}. This is exactly the case in our study, as revealed from modeling results by EUHFORIA. The AR under study was located just slightly above the HCS, which is quite flat. The merged structure shows low tilt with respect to the solar equator and consequently to the HCS. The configuration of the large-scale magnetic field, as given by the HCS, would thus have little relevance for the lateral expansion into EW direction. On the other hand we might expect that its lateral expansion in the NS direction would be affected by the HCS, especially in its south part. Given the weak white-light signatures in the radial direction around the CME apex, it is likely that most of the driving force went into lateral EW expansion, and not radial expansion. The observed type II burst is most likely generated by the shock wave driven with the expanding CME flanks \citep[see e.g.][]{Magdalenic14,2018A&A...615A..89Z}. This is also in agreement with the fact that the interplanetary shock was not observed in\,situ~. The location of the HCS in addition explains the lack of SEPs detected at Earth, as the HCS can serve as a discontinuity boundary making it difficult for charged particles to propagate across. The situation is different for STEREO-A which seems to be situated above the HCS, namely, on the same side of the HCS as the source region of the eruptions under study. This may have favored particles to arrive and be detected at STEREO-A as addressed in Sect.\, \ref{radio}.

In the in\,situ~ signatures a small flux rope (SFR) is observed, most likely associated with the studied solar event with indications of strong interactions with the ambient solar wind plasma and hampered expansion. These can also be related to the fact that the AR and Earth are located at different sides of the HCS. Being located south of the structure, the HCS is likely to hamper its expansion in the south direction and strongly interact and erode its south part. We note that this event falls into a category of so-called \textit{opposite side events} which are also usually related to lower geo-effectiveness \citep[see e.g.,][]{henning85,zhao07}. In addition, as only a partial rotation is observed, it is possible that only one portion of the structure was crossed by the spacecraft, which was already indicated by the simple geometry analysis of the GCS-reconstructed structure's size compared to the relative positions of the AR and spacecraft. Moreover, partial crossing of the spacecraft through the structure, as well as possible erosion are in agreement with its small detected size, as well as the lack of two distinct inner magnetic structures. Finally, the low geomagnetic impact is a direct consequence of the structure's size and magnetic field strength. Although the SFR has south-oriented magnetic field, its strength and duration are too small to produce significant geomagnetic effect \citep[e.g.][]{gonzales94}.

\section{Summary and Conclusion}

We present a detailed analysis of an eruptive event that occurred early on 2019~March~8 in active region AR~12734 (SOL-2019-03-08T03:07C1.3), to which we refer as the International Women's Day event. The event originated from a small AR of moderate magnetic complexity and is characterized by very pronounced eruptive low-coronal signatures, including a small double-peaked C1.3 flare, coronal dimmings, two EUV waves, type II and III bursts, low particle radiation hazard at Earth and two loop ejections evolving into an Earth-directed CME with weak in\,situ~ signatures and negligible Earth-impact.

The two-step nature of the flare energy release profile, as well as the subsequent loop eruptions observed in EUV imagery and the related two-step coronal dimmings indicate two eruptions. These observational findings are supported by the results of the NLFF modelling, showing that there are individually involved portions of the coronal magnetic field over-arched by a single large-scale field. The eruption in the west part of the AR would thus disturb the stability of the east part of the AR, causing the second, sympathetic eruption. However, we find that it is the low-corona magnetic field, bridging a possibly existing low-lying flux rope along the polarity inversion line, which got removed in the two-step flare process, and not the low-lying flux rope (evidenced by an H$\alpha$ filament) itself.

The observation of two EUV waves provides further support for the two-eruption interpretation. The first, globally propagating EUV wave is initiated at the time of the first eruption at the western side of the AR, whereas the second EUV wave (propagating only into the eastern direction) is associated with the eruption at the eastern side of the AR. The first EUV wave was associated with a type II radio burst, indicative of shock formation somewhat higher up in the solar corona, whereas the second event was associated with type III bursts indicative of accelerated electrons moving along newly opened field lines. This is supported by electrons observed within half an hour after the second event at STEREO-A, which had a good magnetic connection to the AR that allowed efficient transport of energetic particles. 

Coronagraphic observations do not unambiguously indicate two separate CMEs, but rather a single entity most likely composed of two sheared and twisted structures that correspond to the two eruptions observed in the low corona. We find that the configuration of the large-scale magnetic field favors the lateral expansion into East-West direction, over the expansion in the North-South direction or radial expansion. Moreover, we find that the CME is propagating on the opposite side of the HCS compared to Earth. The HCS therefore poses an obstacle for both energetic particles and plasma structures, and as such it is very important for interpretation of the in\,situ~ observations. SEPs are not observed at Earth, which may be due to lack of magnetic connectivity. The interplanetary signatures of the solar eruptions are that of a small flux rope with increased density and temperature, most probably related to the strong interaction with the ambient plasma (as would be expected for a structure crossing the HCS). The corresponding geomagnetic impact is practically negligible, as observed in previous cases of \textit{opposite side events}. The analysis of this peculiar event therefore reveals that the large-scale magnetic field can have significant influence on both the early and the interplanetary evolution of the solar eruption.
 
\begin{acknowledgements}
M.D. acknowledges support by the Croatian Science Foundation under the project IP-2020-02-9893 (ICOHOSS). AMV and JKT acknowledge the Austrian Science Fund (FWF): P27292-N20, P31413-N27, I4555. J. G. is supported by the Strategic Priority Program of the Chinese Academy of Sciences (Grant No. XDB41000000) and the National Natural Science Foundation of China (Grant No. 42074222). E.S. is supported by a PhD grant awarded by the Royal Observatory of Belgium. SDO data are courtesy of the NASA/SDO AIA and HMI science teams. The SOHO/LASCO data used here are produced by a consortium of the Naval Research Laboratory (USA), Max-Planck-Institut fuer Aeronomie (Germany), Laboratoire d'Astronomie (France), and the University of Birmingham (UK). SOHO is a project of international cooperation between ESA and NASA. We acknowledge the STEREO/SECCHI consortium for providing the data. The SECCHI data used here were produced by an international consortium of the Naval Research Laboratory (USA), Lockheed Martin Solar and Astrophysics Lab (USA), NASA Goddard Space Flight Center (USA), Rutherford Apple- ton Laboratory (UK), University of Birmingham (UK), Max- Planck-Institute for Solar System Research (Germany), Centre Spatiale de Liege (Belgium), Institut d'Optique Theorique et Appliquee (France), Institut d'Astrophysique Spatiale (France). We acknowledge and thank the ACE SWEPAM and MAG instrument teams and the ACE Science Center for providing the ACE data. The ACE data were obtained from the GSFC/SPDF CDAWeb\footnote{\url{https://cdaweb.gsfc.nasa.gov}}. The Dst used in this paper/presentation was provided by the WDC for Geomagnetism, Kyoto\footnote{\url{http://wdc.kugi.kyoto-u.ac.jp/wdc/Sec3.html}}.
\end{acknowledgements}

\bibliographystyle{aa}
\bibliography{REFs}

\begin{appendix}
\section{NLFF and FV-helicity method description and validation}
\label{app1}
\renewcommand{\theequation}{A\arabic{equation}}
\setcounter{equation}{0}
\renewcommand{\thefigure}{A\arabic{figure}}
\setcounter{figure}{0}

\begin{figure*}
\centering
\includegraphics[width=1.0\textwidth]{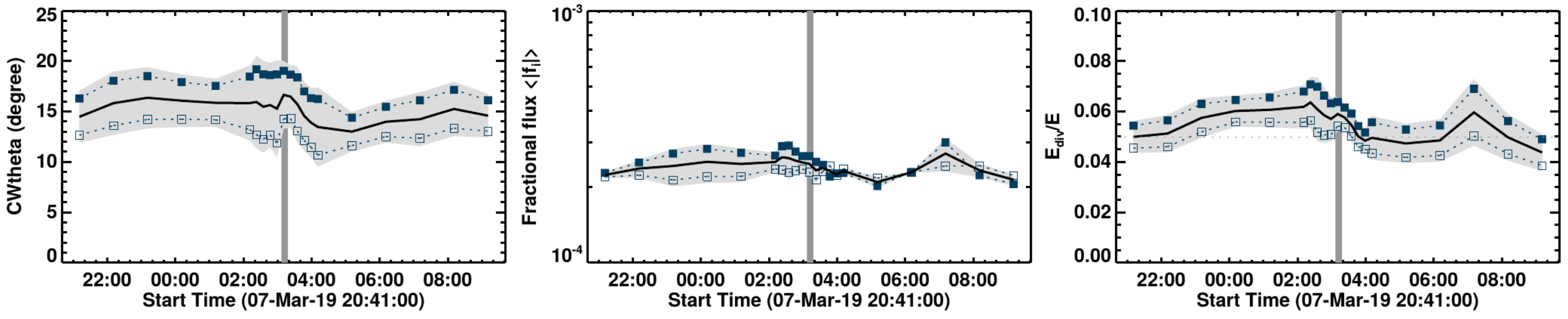}
\put(-490,85){\sf\bf\color{black}(a)}
\put(-315,85){\sf\bf\color{black}(b)}
\put(-145,85){\sf\bf\color{black}(c)}
\caption{Quality of the NLFF model solutions that qualify for analysis, as a function of time. (a) Current-weighted angle between the modeled magnetic field and electric current density. (b) Fractional flux. (c) Non-solenoidal fraction of the total magnetic energy. Filled and empty symbols refer to the preprocessed boundary data being smoothed or non-smoothed, respectively.}
\label{fig_app1}
\end{figure*}

In order to study the coronal magnetic field configuration of AR~12734 in space and time, we use data products from the Helioseismic and Magnetic Imager \citep[][HMI]{2012SoPh..275..229S} and the Atmospheric Imaging Assembly \citep[][ AIA]{2012SoPh..275...17L} on board the Solar Dynamics Observatory \citep[][{\it SDO}]{2012SoPh..275....3P}. We use HMI continuum images images with a cadence of 45~seconds to study the AR's photospheric evolution and, in addition, use AIA images with a time cadence of 24~seconds to investigate the corresponding dynamics at chromospheric/transition region and coronal temperatures. In order to study the three-dimensional (3D) coronal magnetic field, we use {\sc hmi.sharp\_cea\_720s} data which provides the Cylindrical Equal Area (CEA) projected photospheric magnetic field vector within automatically-identified active region  patches \citep[][]{2014SoPh..289.3549B} with the azimuthal component of the vector magnetic field being disambiguated \citep[][]{1994SoPh..155..235M,2009SoPh..260...83L}. We use the full-resolution CEA data (pixel scale $\approx$~360~km at disk center) binned by a factor of two, as an input to a nonlinear force-free (NLFF) method. We use data with the native 12-minute time cadence around the C1.3 flare (between March~08 02:12~UT and 04:12~UT), and use a 1-hour cadence maps covering an extended time period (between March~07 21:12~UT and March~08 09:12~UT). AR~12734 was located close to disk center (N09E01 -- N09W04) during the entire considered time interval (March 21:30~UT  -- 08~March 08:30~UT) so that projection effects can be assumed negligible.

In order to perform the NLFF modeling, we apply the method of \cite{2012SoPh..281...37W}, i.e.\,, we combine the improved optimization scheme of \cite{2010A&A...516A.107W} and a multi-scale approach \citep[][]{2008JGRA..113.3S02W}. In our work, we apply a three-level multi-scale approach to the preprocessed vector magnetic field data. The latter is achieved by applying the preprocessing method of \cite{2006SoPh..233..215W} to the original vector magnetic field data. We adopted a computational domain of $168\times96\times184~\mathrm{pixel}^3$, with the photospheric magnetic flux on the NLFF model lower boundary (at $z=0$) being balanced to within $\sim5\%$. We employ two different free parameter choices for both, the preprocessing and optmization steps, as suggested in \cite{2020A&A...643A.153T}. More precisely, we once run the preprocessing using the standard settings as suggested in \cite{2012SoPh..281...37W} and once without explicit smoothing. Hereafter, we subsequently perform the 3D NLFF reconstruction based on the two time series of preprocessed maps twofold. Once using the equal weightings for the volume-integrated Lorentz force and divergence term \citep[for details see][]{2012SoPh..281...37W}, and one using an enhanced weighting of the volume-integrated divergence to achieve a more solenoidal (divergence-free) solution. Consequently, we obtain four time series of NLFF models, being force- and divergence-free to a different level (which is crucial to be checked for subsequent magnetic helicity analysis; see last paragraph of this Section).

For each of the individual NLFF solution, a potential (reference) field is computed, sufficing $\Bpvec=\nabla\phi$, with $\phi$ being the scalar potential, subject to the constraint $\nabla_n\phi=\Bvec_n$ on $\partial\mathcal{V}$, where $n$ denotes the normal component with respect to the boundaries of $\mathcal V$. For each of the individual NLFF solutions, a potential (reference) field is computed, sufficing $\Bpvec=\nabla\phi$, with $\phi$ being the scalar potential, subject to the constraint $\nabla_n\phi=\Bvec_n$ on $\partial\mathcal{V}$, where $n$ denotes the normal component with respect to the boundaries of $\mathcal V$.

Based on these boundary requirements, and the additional constraint that the NLFF and potential field are solenoidal (i.e.\,, $\nabla\cdot\Bvec=0$ and $\nabla\cdot\Bpvec=0$), we are able to compute the gauge-invariant relative helicity \citep[$\Hv$;][]{1984JFM...147..133B,1985CPPCF...9...111F}, known to represent a meaningful quantity to compute and track the time evolution of a magnetic system within a finite model volume \citep{2012SoPh..278..347V}. In addition, we compute the separately gauge-invariant contribution of the current-carrying field \citep[$\Hj$;][]{2003and..book..345B} in order to gain additional information on the time evolution of the respective non-potential field, $\Bjvec=\Bvec-\Bpvec$.

In order to be able to compute the helicities introduced above, one needs to obtain the vector potentials $\Avec$ and $\Apvec$, which suffice $\nabla\times\Avec=\Bvec$ and $\nabla\times\Apvec=\Bpvec$, respectively.
Therefore, we use the finite-volume (FV) method of \cite{2011SoPh..272..243T} to solve the systems of partial differential equations to obtain $\Avec$ and $\Apvec$, using the Coulomb gauge, $\nabla\cdot\Avec=\nabla\cdot\Apvec=0$. The method has been tested in the framework of an extended proof-of-concept study on FV helicity computation methods \citep[][]{2016SSRv..201..147V}, where it has been shown that for various test setups the methods deliver helicity values in line with each other, differing by only small percentage points. It has also been used in \cite{2019ApJ...880L...6T} to show that the computed helicity is highly dependent on the level to which the underlying NLFF magnetic field satisfies the divergence-free condition.

Successful NLFF modeling, and subsequent helicity computation, requires some numerical conditions to be met. First, the NLFF models are expected to deliver a 3D corona-like model magnetic field with a vanishing Lorentz force and divergence. In order to quantify the force-free consistency we use the current-weighted angle between the modeled magnetic field and electric current density \citep[][]{2006SoPh..235..161S}. In order to quantify the divergence-free consistency of our NLFF solutions we compute the fractional flux metric as introduced by \cite{2000ApJ...540.1150W}. Alternatively, the  non-solenoidal fraction of the total magnetic energy, $\Edivprime$, where $\Ediv$ denotes the energy contribution due to a non-vanishing divergence, has been found to represent a relevant criterion \citep[as first introduced by][]{2013A&A...553A..38V}. In the proof-of-concept study by \citet[][]{2016SSRv..201..147V}, based on solar-like numerical experiments, it was suggested that only $\Edivprime\lesssim0.08$ is sufficient for a reliable helicity computation. In a follow-up study, \cite{2019ApJ...880L...6T} suggested an even lower threshold ($\Edivprime\lesssim0.05$) for solar applications.

Only two of the four employed NLFF model time series suffice the criterion $\Edivprime\leq0.08$ noted above. More precisely, the two time series based on an enhanced weighting of the volume-integrated divergence show a sufficiently low level of $\Edivprime$ during the considered time interval, and are used for further analysis. Having two NLFF model time series of sufficient solenoidal quality at hand (represented by filled and empty symbols in Fig.\,\ref{fig_app1}c, allows us to compute a mean value as well as to indicate a related uncertainty range (black line and gray-shaded area, respectively). For our two NLFF model time series we find a mean value of $\Edivprime\lesssim0.06$. The corresponding mean values describing the force-free and divergence-free consistency are $\theta_j\lesssim15^\circ$ and $<|f_i|>\lesssim3\times10^{-4}$, respectively (see Fig.\,\ref{fig_app1}a and Fig.\,\ref{fig_app1}b, respectively). Having two NLFF model time series of model quality at hand, we are able to compute a mean value of the investigated quantities (magnetic energy ($\Etot$), free magnetic energy ($\Ef$), relative helicity ($\Hv$), and non-potential helicity, $\Hj$) as well as to indicate a related uncertainty.
\end{appendix}

\end{document}